\documentclass[epj]{svjour}
\usepackage{graphics}
\newcommand{\bra}[1]{\langle #1 |}
\newcommand{\ket}[1]{| #1 \rangle}
\newcommand{\beq}{\begin{equation}}
\newcommand{\eeq}{\end{equation}}

\newcommand{\komma}{\;,}
\newcommand{\pkt}{\;.}

\begin{document}

\title{Gross, intermediate and fine structure of nuclear giant resonances: Evidence for doorway states}
\authorrunning{P.~von Neumann-Cosel et al.}
\titlerunning{Gross, intermediate and fine structure of giant resonances}
\author{Peter~von~Neumann-Cosel\inst{1}, Vladimir~Yu.~Ponomarev\inst{1}, Achim Richter\inst{1}, Jochen Wambach\inst{1,2}
}                     
\institute{
Institut f\"{u}r Kernphysik, Technische Universit\"{a}t Darmstadt, D-64289 Darmstadt, Germany
\and
European Centre for Theoretical Studies in Nuclear Physics and related Areas (ECT*) and Fondazione Bruno Kessler, Villa Tambosi, 38123 Villazzano (TN), Italy
}
\date{Received: date / Revised version: date}
%
\abstract{ 
We review the phenomenon of fine structure of nuclear giant resonances and its relation to different resonance decay mechanisms.
Wavelet analysis of the experimental spectra provides quantitative information on the fine structure in terms of characteristic scales.  
A comparable analysis of resonance strength distributions from microscopic approaches incorporating one or several of the resonance decay mechanisms allows conclusions on the source of the fine structure.
For the isoscalar giant quadrupole resonance (ISGQR), spreading through the first step of the doorway mechanism, i.e.\ coupling between one particle-one hole ($1p1h$) and two particle-two hole ($2p2h$) states is identified as the relevant mechanism.
In heavy nuclei it is dominated by coupling to low-lying surface vibrations, while in lighter nuclei stochastic coupling becomes increasingly important.
The fine structure observed for the isovector giant dipole resonance (IVGDR) arises mainly from the fragmentation of the $1p1h$ strength (Landau damping), although some indications for the relevance of the spreading width are also found.  
} 
\maketitle
%


\section{Introduction}
\label{sec1}

\begin{figure}[h]
\begin{center}
\resizebox{0.48\textwidth}{!}{%
  \includegraphics{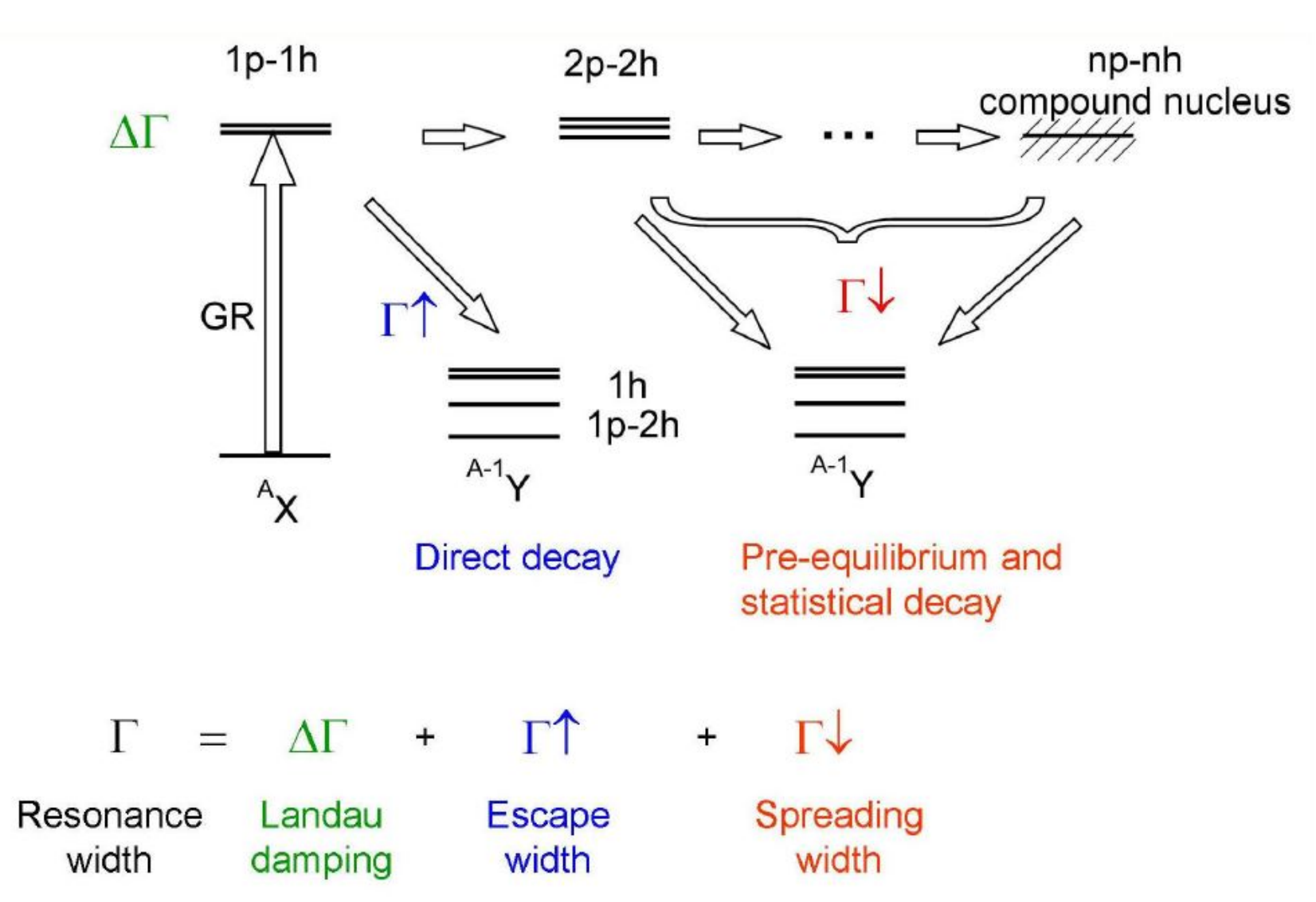}
}
\caption{
Contributions to the decay width of a giant resonance.
For details see text.
Figure taken from ref.~\cite{vnc19}.}
\label{fig11}  
\end{center} 
\end{figure}

Electric and magnetic nuclear giant resonances (GR) are well-known examples of the striking behavior of an interacting system of fermions to form collective modes \cite{spe91,har01}. Over the years, much experimental work has gone into establishing an understanding of the global behavior of their gross features, such as centroid energies and widths. 
It is generally accepted that the total width $\Gamma$ of the resonance is mainly caused by three mechanisms illustrated in fig.~\ref{fig11}: 
fragmentation of the elementary one particle-one hole ($1p1h$) excitations (Landau damping $\Delta \Gamma$)),
direct particle emission from $1p1h$ configurations leading to an escape width $\Gamma^\uparrow$, and the evolution of $1p1h$ configurations into more complicated two-particle two-hole ($2p2h$) and finally to many particle-many hole ($npnh$) states giving rise to a spreading width $\Gamma^\downarrow$
\begin{equation}
\label{eqwidth}
\Gamma = \Delta\Gamma + \Gamma^\uparrow + \Gamma^\downarrow.
\end{equation}

\begin{figure}
\begin{center}
\resizebox{0.48\textwidth}{!}{\includegraphics{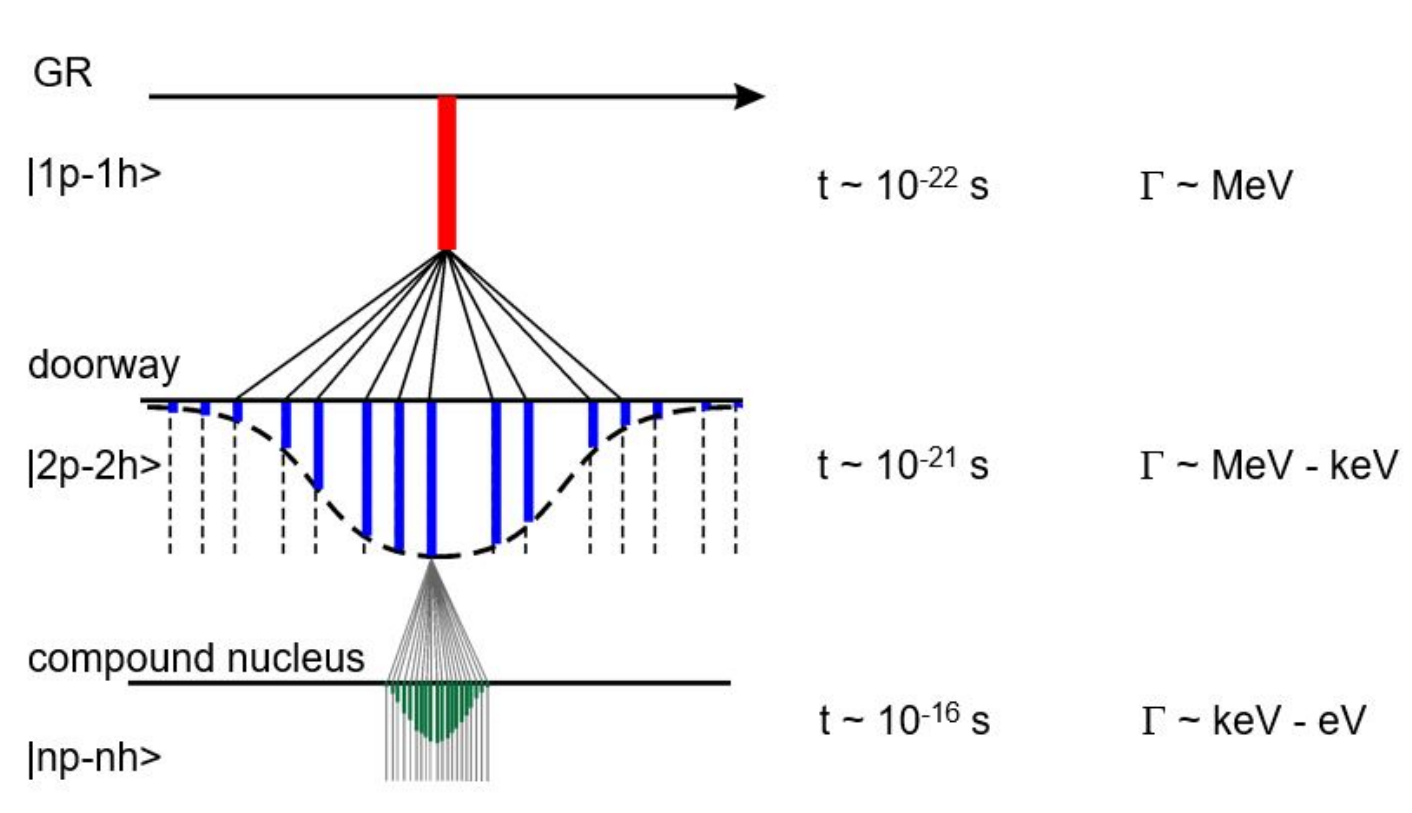}}
\caption{
Doorway state scheme.
For details see text.
}
\label{fig12}  
\end{center} 
\end{figure}

A powerful approach to investigate the role of the different components are coincidence experiments, where direct decay can be identified by the population of 1h and 1p-2h states in the daughter nucleus and the spreading width contribution can be estimated by comparison with statistical model calculations (see, e.g., refs.~\cite{bol88,die94,str00,hun03}).
However, the scheme outlined above also implies a hierarchy of widths and timescales (an assumption underlying all transport theories \cite{cas90,rei94,abe96}) resulting in a fragmentation of the giant resonance strength in a hierarchical manner \cite{bbb98}.
Such a doorway state picture is illustrated in fig.~\ref{fig12} starting from the direct excitation of simple $1p1h$ states. 
The coupling to $2p2h$ states leads to a fragmentation into states acting as "doorways" for the damping of the initial strength across the many complex states until the compound nucleus is reached.
Such a scheme connected to the explanation of intermediate structure has a long history (see, e.g., ref.~\cite{mah69} and references therein).
It implies the existence of lifetimes characteristic for each coupling step with corresponding energy scales ranging from the total width of the order of MeV to the width of compound nuclear states of the order of eV in heavy nuclei.

A challenging theoretical problem is to explain the nature of couplings between the levels in this hierarchy and to predict the scales of the fragmentation of the strength which thus arise from it.
Pier Francesco Bortignon and his collaborators Angela Bracco and Ricardo Broglia at the University of Milano have made important contributions to a solution of this problem \cite{bbb98}.  

Experimentally, it is expected that the coupling scheme leads to fine structure of the giant resonance with characteristic underlying widths, respectively fragmentation patterns.      
While the observation in heavy nuclei has been under debate for a long time \cite{kam97}, systematic studies in the past 15 years have established GR fine structure as a global phenomenon for all types of resonances and across the nuclear chart (see sec.~\ref{subsec21}). 
GR experiments are typically performed using particle beams with energies of several hundred MeV requiring magnetic spectrometers for the detection of scattered particles.
Utilizing beam dispersion matching techniques, energy resolutions of the order of a few tens of keV can be achieved. 
Thus, present-day experimental techniques ( see, e.g., ref.~\cite{vnc19}) are capable to identify characteristic scales that occur between the limits set by the experimental resolution and the broad envelope of the resonances of the order of several MeV. 
By comparison with the scales indicated in fig.~\ref{fig12} it becomes evident that these experiments  should be sensitive to the first coupling step between $1p1h$ and $2p2h$ states.

A variety of methods was proposed to extract quantitative information on scales characterizing the fine structure.   
Early on, an attempt was made to analyze the first high-resolution data on the ISGQR in $^{208}$Pb \cite{sch75,kue81} in terms of a doorway-state model \cite{win83}. 
It could be shown that in this case the spreading width dominates over the escape width but the deduced scales depended strongly on the assumptions about the (unknown) number of doorway states.  
Later, new methods were proposed for the extraction of such scales based an a local scaling dimension approach \cite{aib99,aib03,aib11}, an entropy index method \cite{lac99,lac00}, and the use of wavelet techniques \cite{she04,hei10}.
A comprehensive discussion of the advantages and limitations of the various methods concluded that wavelet analysis is most promising \cite{she08}.

\section{Fine structure of giant resonances}
\label{sec2}

\subsection{Experimental evidence}
\label{subsec21}

The ISGQR was the first case where the fine structure phenomenon was systematically investigated across the nuclear chart \cite{she04,she09,usm11,usm16}.
Interestingly, the fine structure prevails in well-deformed heavy nuclei, where one might expect that the spectral fluctuations are damped  by the extremely high level densities in the ISGQR excitation region. 
Indeed, it was recently shown that rather the fine than the gross structure provides direct evidence for $K$ splitting of the ISGQR in deformed nuclei \cite{kur18}.

\begin{figure}[b]
\begin{center}
\resizebox{0.44\textwidth}{!}{\includegraphics{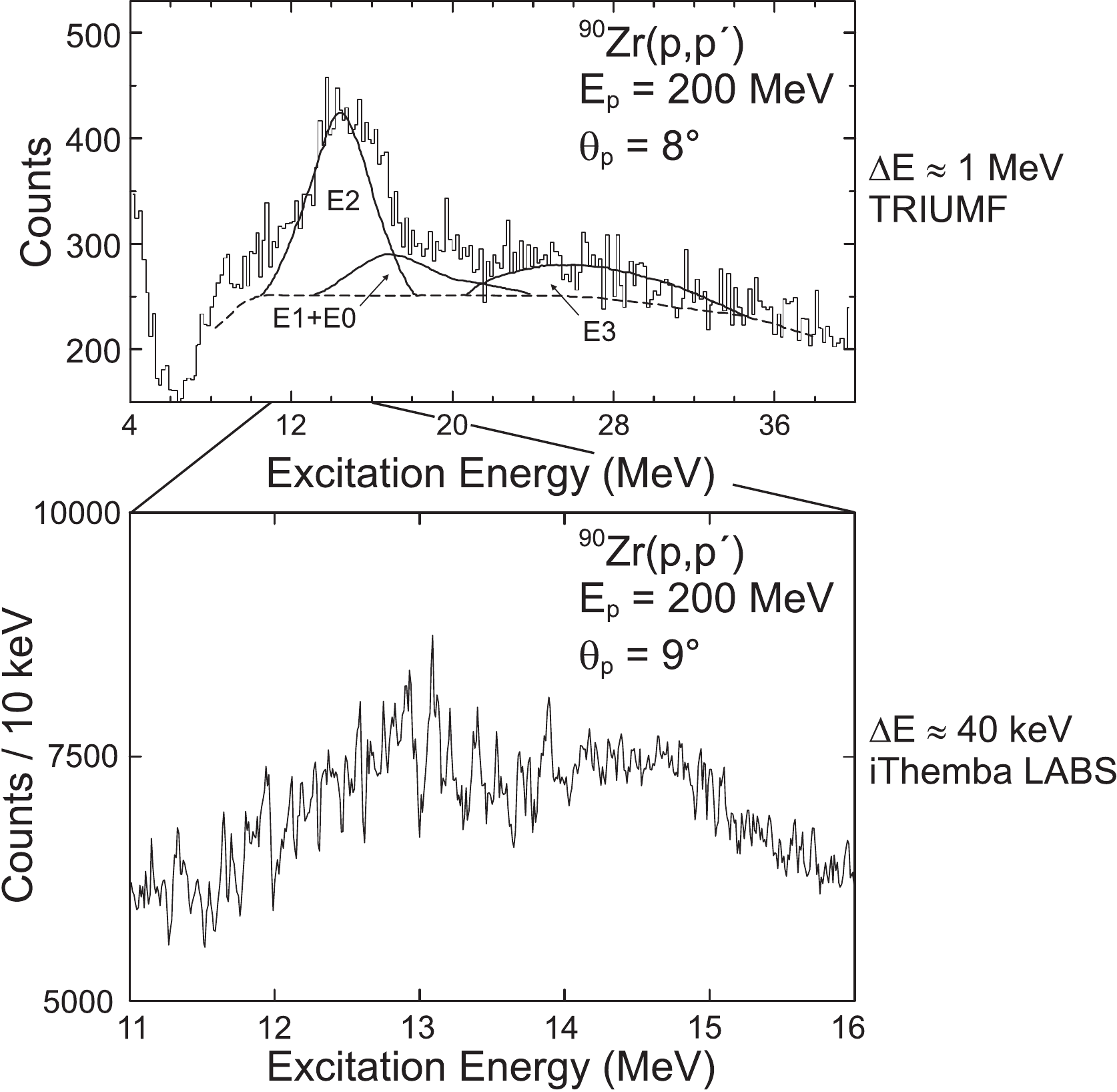}}
\caption{
Comparison of an older study of the ISGQR in $^{90}$Zr with the (p,p$'$) reaction at TRIUMF \cite{ber81} with high-resolution data obtained at iThemba LABS under similar kinematics. 
The broad bump in the upper spectrum around 14~MeV is interpreted as the ISGQR. 
The resolution of $\approx 1$ MeV (FWHM) is insufficient to observe any detailed structure. 
When measured with a resolution $\approx 40$ keV (FWHM), the energy region of the ISGQR expanded in the lower part exhibits fine structure and a double-hump structure deviating from the typical assumption of a single Lorentzian.
Figure taken from ref.~\cite{she09}.
} 
\label{fig211}  
\end{center} 
\end{figure}
The impact of high energy-resolution measurements is illustrated in fig.~\ref{fig211}, in which an early study of the ISGQR in $^{90}$Zr using 200 MeV inelastic proton scattering  at TRIUMF\cite{ber81} is compared to results from a recent measurement at iThemba LABS \cite{she09}, both at a scattering angle where quadrupole transitions are enhanced. 
In ref.~\cite{ber81}, the ISGQR was observed as a broad, smooth, roughly Lorentzian-shaped  ``bump'' at about 14 MeV with a resolution of about 1 MeV (FWHM). 
At the high energy resolution (40 keV FWHM) of the iThemba LABS experiment considerable fine structure is visible. 
In addition, the resonance reveals a double-humped structure deviating from the typical assumption of a single Lorentzian made for the decomposition of hadron scattering spectra as illustrated in the upper part of fig.~\ref{fig211}.

\begin{figure}
\begin{center}
\resizebox{0.35\textwidth}{!}{%
  \includegraphics{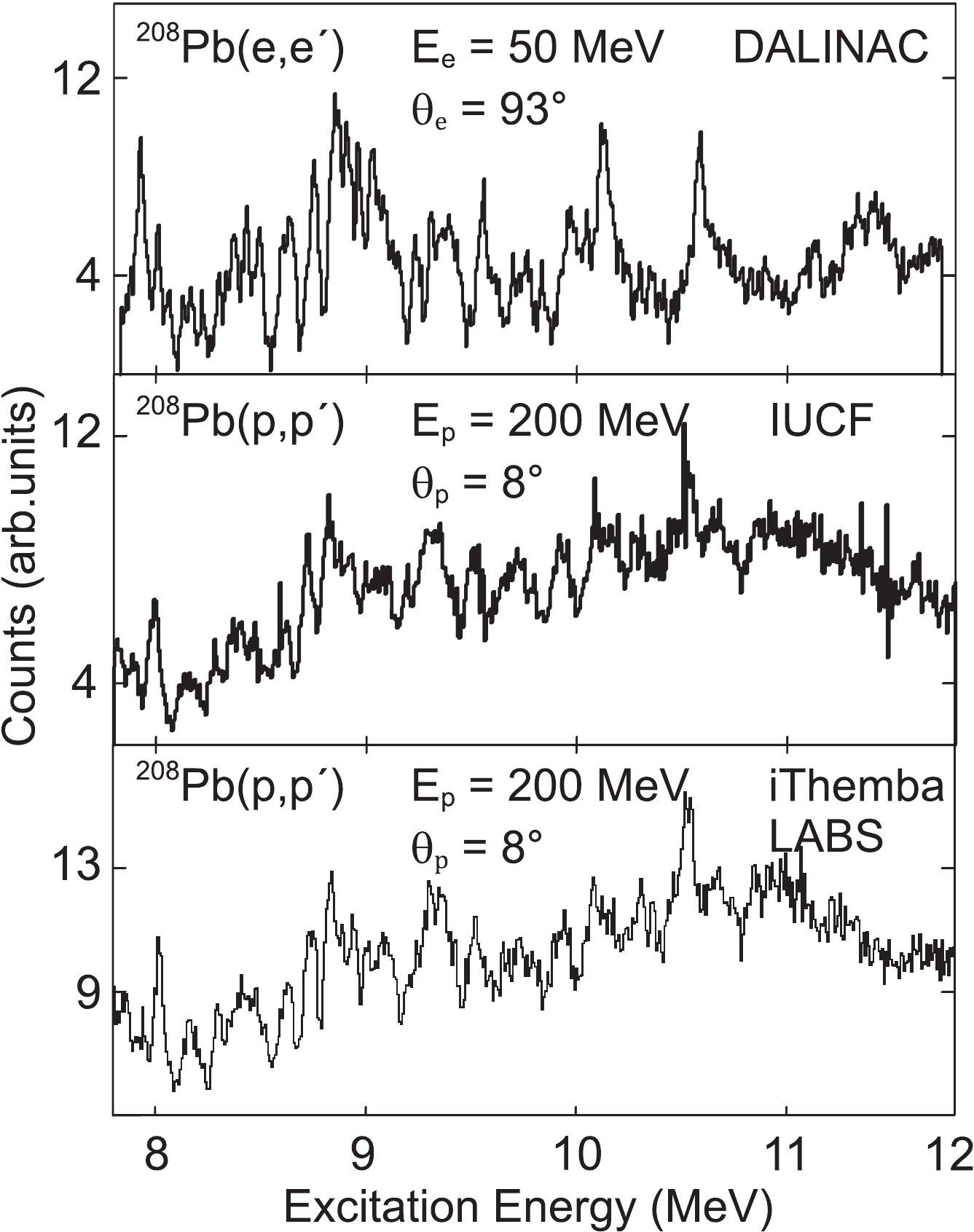}
}
\caption{
Similarity of the structures observed in the three experimental studies of the ISGQR in $^{208}$Pb carried out in Darmstadt (top panel) \cite{kue81}, at
IUCF (middle panel) \cite{kam97}, and at iThemba LABS (bottom panel) \cite{she09}.
Figure taken from ref.~\cite{she09}.
}
\label{fig212}  
\end{center} 
\end{figure}

Figure~\ref{fig212} demonstrates that the observed fluctuations  are indeed of genuine physical origin.
The bottom panel shows a high-resolution $^{208}$Pb(p,p$'$) spectrum measured at iThemba LABS at $\Theta_p=8^\circ$ in the excitation energy region $E_{\rm x} = 8 - 12$ MeV, where the ISGQR is located. 
The middle panel presents a measurement of the same reaction with the same kinematics and comparable energy resolution from IUCF \cite{kam97}. 
The proton scattering data show excellent agreement between the two spectra on a peak-by-peak basis. 
This is also true for high-resolution electron scattering scattering data from the DALINAC \cite{kue81} (top panel), at least up to $E_{\rm x} \simeq 10.5$ MeV. 
 At higher excitation energies some differences between the fine structure in the (e,e$'$) and (p,p$'$) spectra are visible due to the different selectivity of both reactions.
In the electron scattering experiment an excitation of E1 transitions from the low-energy tail of the IVGDR is expected while such transitions are only weakly excited in proton scattering.

\begin{figure}
\begin{center}
\resizebox{0.4\textwidth}{!}{%
  \includegraphics{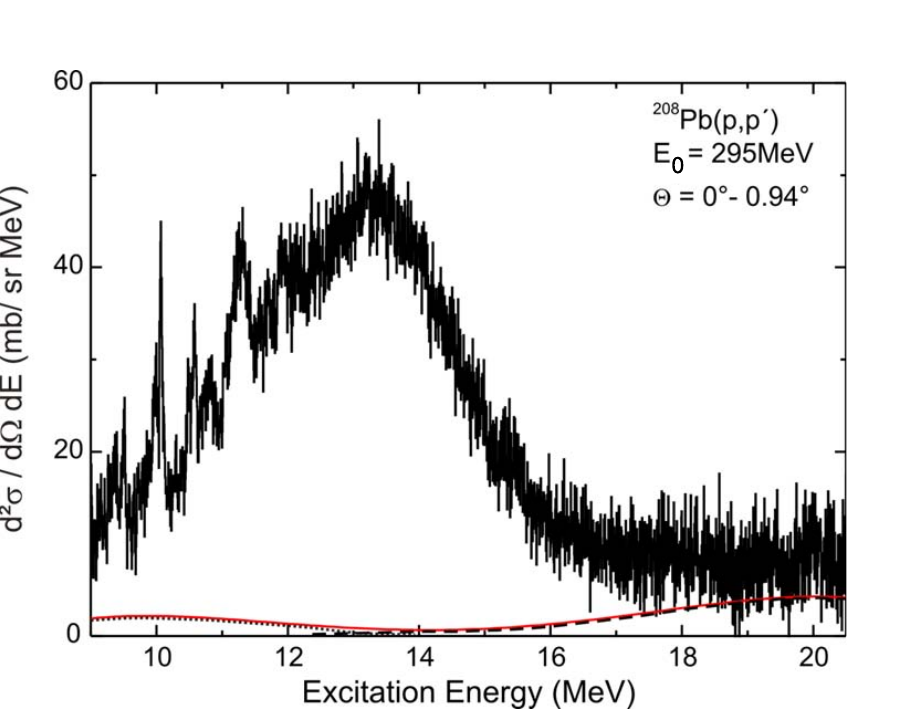}
}
\caption{
Spectrum of the $^{208}$Pb(p,p$'$) reaction at $E_0 = 295$ MeV and $\Theta = 0^\circ - 0.94^\circ$.
The crosss sections reflect E1 excitations induced by relativistic Coulomb excitation.
The background from non-E1 excitations (full line) is determined by a MDA with contributions from excitation of E2 strength (dotted line) and a phenomenological component (dashed line) \cite{pol12}.
Figure taken from ref.~\cite{pol14}.
}
\label{fig213}  
\end{center} 
\end{figure}

In the last years, the fine structure phenomenon has been experimentally established for the IVGDR as well. \cite{tam11,iwa12,kru15,don18,fea18,jin18}.   
This has been facilitated by the realization of inelastic proton scattering experiments at energies of a few hundred MeV under extreme forward angles including $0^\circ$ combined with high energy resolution achieved by dispersion matching techniques \cite{vnc19}.
In these kinematics the cross sections are dominated by relativistic Coulomb excitation populating the IVGDR. 
Figure~\ref{fig213} shows a spectrum of the $^{208}$Pb$(p,p')$ reaction at $E_0 = 295$ MeV and covering an angular range $\Theta = 0^\circ - 0.94^\circ$.
The full (red) line indicates the background from other contributions to the spectrum deduced by a multipole decomposition analysis (MDA) \cite{tam11,pol12}.
The main contributions are from excitation of the ISGQR and to a lesser extent from the ISGMR (dotted line) and a phenomenological part (dashed line) dominated by quasifree reactions.
In any case, the contributions under the IVGDR peak are small justifying the assumption that they do not influence the fluctuations visible in the data. 
The cross section fluctuations are particularly pronounced on the lower side of the IVGDR and are damped on the upper side, most likely due to the strong increase of the level density of $J^\pi = 1^-$ states across the resonance.  

\begin{figure}
\begin{center}
\resizebox{0.48\textwidth}{!}{%
  \includegraphics{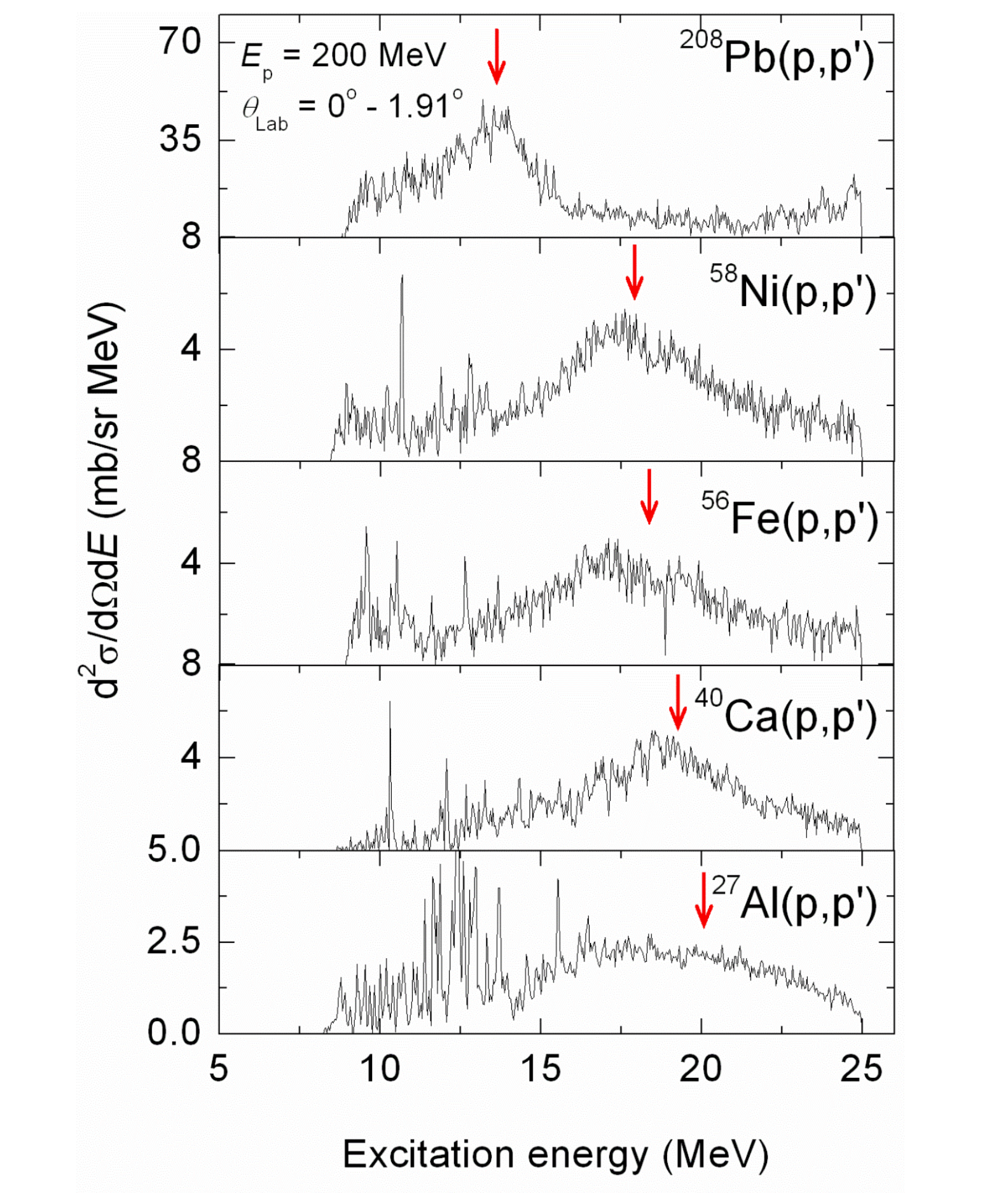}
}
\caption{
Experimental double differential cross sections for the  spectra (from top to bottom) of the $^{208}$Pb, $^{58}$Ni, $^{56}$Fe ,$^{40}$Ca and $^{27}$Al(p,p$'$) reactions at $E_{\rm p}$ = 200 MeV and scattering angles $\theta_{\rm lab} = 0^{\circ} - 1.91^{\circ}$. 
The red arrows show the centroids of the IVGDR expected from systematics.
Figure taken from ref.~\cite{jin18}.
}
\label{fig214}  
\end{center} 
\end{figure}

Figure~\ref{fig214} presents results from a recent survey performed at $E_0$ = 200 MeV at iThemba LABS \cite{jin18}.
The maximum of the prominent bump visible in all data follows the systematics of the IVGDR \cite{har01} 
\begin{equation}
E_{\rm C} = 31.2 \, A^{-1/3} + 20.6 \, A^{-1/6},
\label{eq:gdrec}
\end{equation}
except for $^{27}$Al.
However, the IVGDR in light-mass nuclei is known to be extremely fragmented and to extend to very high energies \cite{era86} such that part of the E1 strength is likely to be outside the momentum acceptance of the spectrometer. 
Pronounced fine structure is visible over the excitation energy region of the IVGDR in all nuclei investigated, thus confirming the global character of this phenomenon.
The overall cross sections decrease considerably with atomic number indicating that they are dominated by Coulomb excitation.

Finally, we note that fine structure has also been established in magnetic and spinflip resonances. 
High-resolution electron scattering studies in Darmstadt demonstrate the fragmentation of magnetic dipole \cite{met87,lan04} and quadrupole \cite{fre78,meu80,meu81,vnc99,rei02} modes. 
The Gamow-Teller resonance in heavy nuclei studied with high resolution utilizing the ($^{3}$He,t) reaction at RCNP again shows pronounced fine structure  \cite{kal06,fre18}.

\subsection{Quantitative analysis of the spectra using wavelet techniques}
\label{subsec22}

\begin{figure}
\begin{center}
\resizebox{0.48\textwidth}{!}{%
  \includegraphics{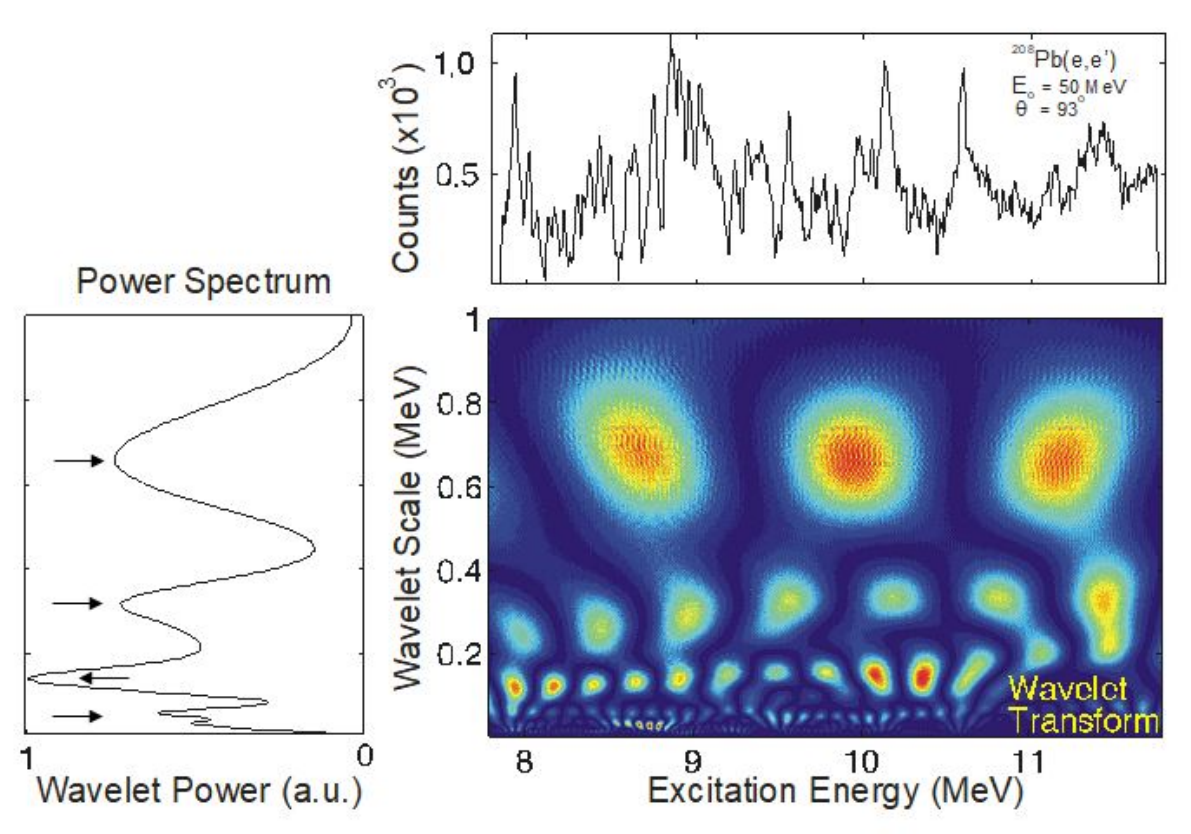}
}
\caption{
Top right: Spectrum of the $^{208}$Pb (e,e$^\prime$) reaction \cite{kue81}. 
Bottom right: Squares of the wavelet coefficients, Eq.~(\ref{eq:wc}), as a function of excitation energy from a CWT. 
Bottom left: Projection of the wavelet coefficients on the scale axis (power spectrum).
The arrows indicate characteristic scales.
}
\label{fig221}  
\end{center} 
\end{figure}

As pointed out in the introduction, a variety of methods have been proposed for a quantitative analysis of the fine structure in the spectra.
These have been compared with each other in ref.~\cite{she08} and the method of wavelet analysis has been identified as the most promising approach. 
The wavelet analysis of the measured spectra is illustrated by the example of  ISGQR data from the the $^{208}$Pb(e,e$^\prime$) reaction (fig.~\ref{fig221}). 
It proceeds via the calculation of a wavelet coefficient $C$ from the measured cross sections $\sigma(E)$ (expressed here in counts/channel) shown in the top right part 
\begin{equation}
C_i(\delta E)\equiv C(\delta E,E_i)=\frac{1}{\sqrt{\delta E}} \int \sigma(E) \Psi^*\left(\frac{E_i-E}{\delta E}\right) dE,
\label{eq:wc}
\end{equation}
where $E_i$ is the excitation energy of channel $i$, $\delta E$ the wavelet scale, and $\Psi$ the wavelet function. 
Here, the complex Morlet wavelet 
\begin{equation}
\Psi(x)=\pi^{-1/4}\,e^{ik_0x}\,e^{-x^2/2},
\label{eq:wf}
\end{equation}
with $k_0=5$ is employed, which provides optimum balance between resolution of excitation energy and energy scale for the present application (see, e.g., also ref.~\cite{usm16}).
The wavelet decomposition is done over the whole spectrum with reflective boundary conditions to avoid finite-range-of-data errors.  
The analysis of the fine structure of giant resonances is performed using the continuous wavelet transform (CWT), where the fit procedure can be adjusted freely to the required precision.
Further details can be found e.g.\ in refs.~\cite{she08,pol14,usm11,usm16,fea18}. 

The squares of the wavelet coefficients, representing a measure of the strength of structures resolved by the wavelets, are displayed in the bottom-right part of fig.~\ref{fig221}.
The colour code indicates their magnitude from red (large) to blue (small).
At certain wavelet scale values maxima are observed across the ISGQR excitation region.
The structure of neighboring maxima/minima along these lines results from the oscillating form of the wavelet function.
It is convenient to project  the two-dimensional distribution on the scale axis.
The resulting power spectrum
\begin{equation}
P_w(\delta E)=\frac1N \sum_{i=i_1}^{i_2}|C_i(\delta E)C^*_i(\delta E)|,
\label{eq:power}
\end{equation}
where $i_1$ and $i_2$ indicate the boundaries of the region of interest, is shown in the bottom left part of fig.~\ref{fig221}.
Peaks of strength in this power spectrum are associated with characteristic scales of the structures in the region of the ISGQR. 
The power spectrum is normalized to the spectral variance in order to facilitate comparison between different nuclei and with theoretical results. 
The analysis of the fluctuations, if represented as a power, characterizes the variance of the series under consideration. 
The Fourier transform preserves the variance of the signal and the CWT does as well (at least approximately) since it is a convolution. 
Thus, a normalization to the variance facilitates a comparison of powers deduced from the various spectra despite the absence of an absolute scale.

\section{Theoretical approaches}
\label{sec3}

While the wavelet analysis described in sec.~\ref{subsec22} provides a quantitative measure of the fine structure, there is no direct way to relate these characteristic scales to the decay mechanisms illustrated in fig.~\ref{fig11}.
The wavelet coefficients depend on underlying widths as indicated in the doorway scheme of fig.~\ref{fig12}, but also expected from the escape width.
Furthermore, they are influenced by characteristic energy differences between peaks of the fine structure, which can be induced by Landau damping or by the coupling to complex states, and there is no way to decompose these different contributions to the wavelet coefficients.    
Thus, any interpretation of the characteristic scales requires a corresponding analysis of theoretical spectra from models which include some or all of the giant resonance decay mechanisms.

\subsection{General considerations}

Many-body methods for describing collective modes of medium and heavy nuclei are based on the notion of a common mean field (spherical or deformed) in which -- in the absence of pair correlations --
nucleons move independently in shells that are filled up to the respective Fermi levels for protons and neutrons. The nuclear Hamiltonian for A nucleons, $\hat H_A$, can therefore be expressed in second quantized form in terms of fermionic creation and annihilation operators, $a^\dagger_i$ and $a_i$, involving the quantum numbers of the $i$-th orbital: 
\beq
\hat H_A=\sum_i\epsilon_ia^\dagger_ia_i+\frac{1}{4}\sum_{ijkl}v_{ijlk}a^\dagger_ia^\dagger_ja_la_k\pkt
\label{HA}
\eeq
Here, $\varepsilon_i$ denotes the single-particle energies in the mean field and $v_{ijkl}$ the antisymmetrized matrix elemenents of an effective two-body interaction
ignoring three- and higher-body interactions.
The mean field is either taken phenomenologically or determined selfconsistently from the effective two-body interaction in the Hartree-Fock (HF) approximation. In principle, the effective interaction has to be obtained from the underlying bare nucleon-nucleon interaction but in applications is often parametrized. 

Given the exact eigenstates $\ket{\nu}$
of the many-body Schr\"odinger equation
\beq
\hat H_a\ket{\nu}=E_\nu\ket{\nu}\komma
\label{MBSE}
\eeq
where $\nu$ denotes states of total angular momentum $(J,M)$ and parity $\pi$,
the nuclear excitation spectrum in response to a weak external perturbation 
\beq
\hat F(t)=\hat Fe^{i\omega t}+\hat F^\dagger e^{-i\omega t}
\eeq
is given by the strength function $S_F(\omega)$
\beq
S_F(\omega)=\sum_\nu|\bra{\nu}\hat F\ket{0}|^2\delta(\omega-E_\nu)\komma
\label{SFexact}
\eeq
which is of second order in the perturbation.

Sum rules are energy-weighted moments of the strength function $S_F(\omega)$:
\beq
m_F^k=\int d\omega \omega^kS_F(\omega),\quad k=0\pm1,\pm2,\cdots
\eeq
that can be evaluated  as ground-state expectation values and allow for a simplified treatment of the average propetries of collective motion. They are also related to bulk constants of the system, such as static polarizabilities and susceptibilities. One may define a set of energies
\beq
{\cal E}^k_F=(m^k_F/m^{k-2}_F)^{1/2}\komma
\eeq
which characterize the strength distribution. If it is sharply peaked at a certain energy, then all ${\cal E}^k_F$ essentially coincide. The degree to which they differ reflects the width, asymmetry etc.\ of the distribution. 

For medium-heavy and heavy nuclei the exact solution of the many-body Schr\"odinger equation (\ref{MBSE}) is numerically prohibitive and appropriate approximation schemes have to be developed. These are based on the fact that the elementary excitations on a mean-field groundstate are of particle-hole nature and a expansion of the many-body wavefunctions in terms of p-h operators,
or in the case of superfluidity pairs of BCS quasiparticle operators,
is the natural starting point. To lowest order this leads to the Random-Phase-Approximation (RPA or QRPA) which can be formulated both non-relativistically and relativistically. Higher orders in the expansion lead to extended RPA models such as the Second Random-Phase-Approximation (SRPA) or (Quasi)Particle-Phonon Coupling models.       

The equations  for extended RPA models start from a formal creation operator $\hat Q_\nu^\dagger$ of an exact eigenstates of the many-body Schr\"odinger equation (\ref{MBSE}) defined through
\beq
\ket{\nu}=\hat Q_\nu^\dagger\ket{0},\quad \hat Q_\nu\ket{0}=0\quad  {\rm for\; all}\;\;\nu  
\eeq
\beq
[\hat H_A,\hat Q_\nu^\dagger]\ket{0}=(E_\nu-E_0)\ket{0}\pkt
\eeq
Here $\ket{0}$ and $E_0$ denote the exact ground state vector and its energy, respectively. 
The operator $\hat Q_\nu^\dagger$ is now expanded in a set of bosonic operators $\{\hat O^\dagger_i\}$ composed of a set $\{\hat b^\dagger_j\}$  and its hermitian conjugate $\{\hat b_j\}$ as
\beq
\hat Q_\nu^\dagger=\sum_iP^\nu_i\hat O^\dagger_i=\sum_j\left(X^\nu_j\hat b^\dagger_j-Y^\nu_j\hat b_j\right)\pkt
\label{Qspace}
\eeq   
Multiplying from the left by an arbitrary variation $\bra{0}\delta\hat O$ on the exited-state vector
$\bra{\nu}$ and setting $E_0=0$ yields
\beq
\bra{0}[\delta\hat O,[\hat H_A,\hat Q_\nu^\dagger]]\ket{0}=E_\nu\bra{0}[\delta\hat O,\hat Q_\nu^\dagger]\ket{0}\komma
\label{EoM}
\eeq
whose solutions are equivalent to those of the original Schr\"odinger equation.

\subsection{Random-Phase-Approximation (RPA)}

As the exact problem cannot be solved in practice the operator set $\{\hat b_j\}$ and$\{b_j\}$ has to be restricted. As noted above the mean-field picture of independent particle motion suggests to start by approximating $\{\hat b^\dagger_j\}$  and  $\{\hat b_j\}$  by fermion pair operators for $1p1h$ excitations:
\beq
\hat b^\dagger_1\equiv a^\dagger_{p}a_h;\quad \hat b_1\equiv a^\dagger_ha_p 
\eeq  
such that
\beq
\hat Q_\nu^\dagger=\sum_1\left(X^\nu_1\hat b^\dagger_1-Y^\nu_1\hat b_1\right)=\sum_{ph}\left(X^\nu_{ph}a^\dagger_pa_h-Y^\nu_{ph}a^\dagger_ha_p\right)
\label{RPAw}
\eeq
Restricting the variation $\delta O$ to the p-h subspace and replacing the expectation values of the commutators in eq.~(\ref{EoM}) by a Slater determinant of independent particles (HF-groundstate), the equations of motion  lead to the RPA equations
\beq
\left (\begin{array}{cc}
A_{11'}& B_{11'}\\
-B^*_{11'}& -A^*_{11'}
\end{array}\right )
\left (\begin{array}{c}
X^\nu_{1'}\\
Y^\nu_{1'}
\end{array}\right )=
E^\nu_1
\left (\begin{array}{c}
X^\nu_{1}\\
Y^\nu_{1}
\end{array}\right )
\label{RPAeq}
\eeq
where 
\beq
A_{11'}=\varepsilon_{ph}\delta_{pp'}\delta_{hh'}+v_{ph'hp'};\quad B_{1,1'}=v_{pp'hh'}
\eeq
and $E^\nu_1$ are the RPA values for the excitation energies.

For external perturbations of one-body type 
\beq
\hat F=\sum_{ij}f_{ij}a^\dagger_ia_j
\label{Fonebody}
\eeq
{\it i.e.} the external probe (electromagnetic, hadronic or weak) couples to a single nucleon the strength function  $S_F(\omega)$ can be recast in the following form: 
\beq
S_F(\omega)=-\frac{1}{\pi}\sum_{11'}f^*_1R_{11'}(\omega)f_{1'}
\eeq  
where $R_{11'}$ denotes the $ph$ propagator and thus the indices $1$ and $1'$ are either of p-h or h-p type. In terms of the RPA-matrices $A_{11'}$ and $A_{11'}$ $R_{11'}$ is given by  
\beq
R_{11'}(\omega)=\left (\begin{array}{cc}
\omega+i\eta-A_{11'}&B_{11'}\\
-B^*_{11'}&-\omega+i\eta-A^*_{11'}
\end{array}\right )^{-1}\pkt
\eeq
The coherent sum (\ref{RPAw}) leads to collective modes in the excitation spectrum, most prominently Giant Resonances, which exhaust a large part of the energy-weighted sum rule $m^1_F$.The energy-weighted sum rule ($f$-sum rule) for oscillations associated with a conserved current is the same as that for non-interacting fermions. This is a crucial test of the consistency of a given approximation with general conservation laws imposed by symmetry. The RPA obeys the $f$-sum rule. The RPA also describes the fragmentation of collective strength into individual p-h excitations $\Delta\Gamma$ which is commonly referred to as "Landau damping" and the escape of single nucleons $\Gamma^\uparrow$ by a proper treatment of the continuum.  
 
The mean-field approach, underlying Eq.~(\ref{HA}), forms the basis for several improvements beyond the (Q)RPA. 
Since they are referred to in comparisons with experiment below, we mention the Extended Theory of Finite Fermi Systems (ETFFS) \cite{Kam} and the Relativistic Quasiparticle Time-Blocking Approximation (RQTBA) \cite{RQTBA} in which simple $1p1h$ configurations are coupled to complex configurations of the type: $1p1h \otimes \textrm{phonon}$ or $2qp \otimes \textrm{phonon}$ for open shell nuclei.  
Low-energy phonons of several multipolarities are usually involved. 
Their energies and amplitudes are obtained by solving the QRPA equations.
The Wood-Saxon potential for the mean-field and the residual Landau-Migdal
interaction are used in the ETFFS (see also \cite{Lyuto} for a self-consistent treatment). 
The RQTBA is a self-consistent approach and is based on covariant energy-density functionals. Another method we will refer to is based on extensions of  the time-dependent Hartree-Fock (ETDHF) theory \cite{ETDHF}. 
Here one divides the space of complex configurations into two sub-spaces: the sub-space of $2p2h$ configurations which represents the incoherent dissipation mechanism due to nucleon-nucleon collisions and the sub-space of $1p1h \otimes {\rm phonon}$ configurations responsible for the coherent dissipation mechanism. 
Effective Skyrme forces are used in calculations by this model.

More explicitly we present in the following two approaches. One is the second SRPA whose derivation, in line with the previous formal discussions is particularly transparent. The other is the Quasiparticle-Phonon Model (QPM) \cite{Solo,LoIu}, which is used  extensively in comparisons with experiment as described below.

\subsection{Second Random-Phase-Approximation (SRPA)}
\label{subsec31}

As an obvious improvement of the RPA one may enlarge the operator space $\{\hat O^\dagger_i\}$ in 
Eq.~ (\ref{Qspace}) to include also $2p2h$ excitations with
\beq
\hat b^\dagger_2\equiv a^\dagger_{p_1}a^\dagger_{p_2}a_{h_1}a_{h_2};\quad
 \hat b_2\equiv a^\dagger_{h_1}a^\dagger_{h_2}a_{p_2}a_{p_1} 
\eeq 
such that
\beq
\hat Q_\nu^\dagger=\sum_1\left(X^\nu_1\hat b^\dagger_1-Y^\nu_1\hat b_1\right)
+\sum_2\left(X^\nu_2\hat b^\dagger_2-Y^\nu_2\hat b_2\right)\pkt
\eeq
Following steps similar to those for deriving the RPA equations from Eq.~(\ref{EoM}) one arrives at the SRPA equations:
\beq
\left (\begin{array}{cccc}
A_{11'}& B_{11'}&A_{12'}&0\\
-B^*_{11'}& -A^*_{11'}&0&-A^*_{12'}\\
A_{21'}&0&A_{22'}&0\\
0&-A^*_{21'}&0&-A^*_{22'}
\end{array}\right )
\left (\begin{array}{c}
X^\nu_{1'}\\
Y^\nu_{1'}\\
X^\nu_{2'}\\
Y^\nu_{2'}
\end{array}\right )=
E^\nu_2
\left (\begin{array}{c}
X^\nu_{1}\\
Y^\nu_{1}\\
X^\nu_{2'}\\
Y^\nu_{2'}\\
\end{array}\right )
\eeq
which now involve $1p1h$  as well as $2p2h$ excitations and their couplings. The explicit form of the matrix elements $A_{12'}$ and $A_{22'}$ can be found e.g.\ in ref.~\cite{dro90}. On its diagonal $A_{22'}$ contains the unperturbed $2p2h$ energies $\varepsilon_{p_1h_1}+ \varepsilon_{p_1h_1}$  Obviously the matrix dimension is much larger than that of the RPA and, for heavy nuclei, requires major computational efforts for their solutions.

Again restricting oneself the external perturbations $\hat F$ of one-body type (\ref{Fonebody}) the SRPA strength function $S_F$ is again given by the $ph$ propagator
\beq
S_F(\omega)=-\frac{1}{\pi}\sum_{11'}f^*_1R_{11'}(\omega)f_{1'}
\eeq  
that now includes the coupling to $2p2h$ states 
\beq
R_{11'}(\omega)=\left (\begin{array}{cc}
\omega+i\eta-\tilde A_{11'}(\omega)&B_{11'}\\
-B^*_{11'}&-\omega+i\eta-\tilde A^*_{11'}(-\omega)
\end{array}\right )^{-1}\pkt
\eeq
This looks exactly like the RPA propagator except that the A-matrix is now frequency dependent and acquires a complex contribution to the coupling to $2p2h$ excitations
\beq
\tilde A_{11'}(\omega)=A_{11'}+\sum_{22'} A_{12}\left(\omega+i\eta -A_{22'}\right)^{-1}A_{2'1'}\pkt
\eeq
Physically this corresponds to the "spreading width" $\Gamma^\downarrow$
\beq
\Gamma^\downarrow_{11'}(\omega)=-2\;{\rm Im}\;\tilde A_{11'}(\omega)
\eeq
which is a matrix in $1p1h$ space.

It can be shown that the non-energy weighted and the energy-weighted sum rules $m^0_F$ and $m^1_F$ are the same in RPA and SRPA. This implies that $2p2h$ states redistribute the strenght without creating additional strength as compared to the RPA\footnote{This is a consequence of the one-body nature of the external field $F$. If it involves two-body operators arising e.g.\ from "exchange currents" this is no longer the case.}. It is particularly satisfying to see that the SRPA meets also fulfils the $f$-sum rule as required by symmetry arguments. 

In passing we note that in ref.~\cite{vas18} a systematic study of the ISGQR has been performed with a so-called subtracted SRPA (SSRPA)  model in the framework of an energy density functional approach.

\subsection{Quasiparticle-Phonon model (QPM)}
\label{subsec32}

The SRPA involves the diagonalization of the nuclear Hamiltonian in the basis of $1p1h$ and $2p2h$ states, which poses severe numerical limitations for heavy nuclei. Instead the Quasiparticle-Phonon model is based on the doorway concept as discussed in detail in \cite{bbb98}. In heavy nuclei it is numerically much more tractable and, in particular, allows to also include $3p3h$ excitations.
In the QPM, the information on physical observables
is carried by simple doorway states of the $1p1h$ (two-quasiparticle) nature which are 
embedded in a high density of background states that have more complex 
structure and cannot be easily excited from the ground state in nuclear 
reactions. The interaction between the doorway and background states
leads to a fragmentation of the physical strength over many states of a mixed 
($1p1h + 2p2h + \ldots$) nature as presented in fig.~\ref{fig12}.

The QPM is implemented by means of the step-by-step diagonalization of a given  
model Hamiltonian (Eq.~(\ref{HA})). From the onset, pair-correlations of a possible superfluid ground 
state in open-shell nuclei are included. Thus in a first step, the BCS equations are 
solved including monopole pairing interactions in the second term of Eq.~(\ref{HA}). In BCS theory
a set of quasiparticle operators $\alpha_i^{\dagger}$ is introduced as a linear combination of fermion creation and annihilation operators $a^\dagger_i$ and $a_i$ as
\beq
\alpha_i^{\dagger} = u_i a_i^{\dagger} + v_i a_i
\label{BogoT}
\eeq
where $u_i$ and $v_i$ are the coefficient of the Bogoliubov transformation.

Then, the QRPA equations, similar to (\ref{RPAeq}), are solved, i.e. 
the Hamiltonian is diagonalized on the basis of two-quasiparticle states. In terms of the quasiparticle operators (\ref{BogoT}) it involves parts of the residual interaction in Eq.~(\ref{HA}) proportional to 
$\alpha_i^{\dagger} \alpha_j^{\dagger} \alpha_l^{\dagger} \alpha_k^{\dagger}$,
$\alpha_i^{\dagger} \alpha_j^{\dagger} \alpha_l \alpha_k$ and
$\alpha_i \alpha_j \alpha_l \alpha_k$.
The solutions of the the QRPA equations provide a set "phonon" states in the way similar to Eq.~(\ref{RPAw}) for different values ($J^\pi$) and thus yield a set $i=1, 2, \ldots$  of natural-parity ($1^-_i$, $2^+_i$, $\ldots$) and unnatural-parity ($1^+_i$, $2^-_i$, $\ldots$) phonons. 

The QRPA one-phonon states, with their predominant $1p1h$ nature are strongly excited from the 
ground state (the phonon vacuum or $0p0h$ state) by the 
one-body external field (\ref{Fonebody}) and thus, have features of doorway states.

In a next step the wavefunctions of the excited states $\ket{\nu}\equiv\ket{J^\pi,i}$ are expanded in QRPA one-phonon states. Up to third order, which includes $1p1h, 2p2h$ and $3p3h$ excitations, one has 
\begin{eqnarray}
\ket{\nu} =    
\left\{ \sum_i R^{Ji}\hat Q^\dagger_{J, i}  
+
\sum_{J_1 i_1, J_2 i_2 }  
P^{J_2 i_2}_{J_1 i_1} 
[\hat Q^\dagger_{J_1 i_1} \hat Q^\dagger_{J_2 i_2}]_J 
\right .  
\nonumber\\  
+
\left .  
\sum_{J_1 i_1, J_2 i_2, J_3 i_3 } 
T^{J_3 i_3}_{J_1 i_1 J_2 i_2} 
[ [\hat Q^\dagger_{J_1 i_1} \hat Q^\dagger_{J_2 i_2}] \hat Q^\dagger_{J_3 i_3}]_{J} 
\right \} \ket{0}\pkt
\nonumber\\
\end{eqnarray}
Notice, that complex configurations of the states $\nu$ of angular momentum $J$
may be built up of phonons of different angular momenta
$J_1, J_2, J_3 \neq J$.
 
To obtain the expansion coefficients $R^\nu, P^\nu$ and $T^\nu$ one has to solve
\begin{equation}
\left(
\begin{array}{ccc}
E_{1ph} & V_{1ph}^{2ph} & 0 \\
V_{1ph}^{2ph} & E_{2ph} & V_{2ph}^{3ph} \\
0 & V_{2ph}^{3ph} & E_{3ph} 
\end{array}
\right)
\left(
\begin{array}{c}
R^{\nu}\\
P^{\nu}\\
T^{\nu}
\end{array}
\right)
=
E^{\nu}
\left(
\begin{array}{c}
R^{\nu}\\
P^{\nu}\\
T^{\nu}
\end{array}
\right)
\label{QPM}
\end{equation}
which accounts for the interaction between 
doorway and background states and involves the
$\alpha_i^{\dagger} \alpha_j^{\dagger} \alpha_l^{\dagger} \alpha_k$ and
$\alpha_i^{\dagger} \alpha_j \alpha_l \alpha_k$
terms of the residual interaction in Eq.~(\ref{HA}).
The matrix $E_{1ph}$ contains the energies of 
one-phonon states, $E_{2ph}$ -- the energies of two-phonon states, 
etc; and the matrix $V_{1ph}^{2ph}$  comprises the matrix elements of 
the interaction between one- and two-phonon configurations, etc.

The matrix $E_{1ph}$ has a diagonal form because the QPM Hamiltonian 
is already pre-diagonalized in the QRPA. The matrices $E_{2ph}$ and $E_{3ph}$ 
are also diagonal 
when phonons are considered as bosons. Accounting for the internal
fermionic structure of the phonons leads to interaction between two- and two- 
(three- and three-, etc) phonon configurations; then the matrices
$E_{2ph}$ and $E_{3ph}$ are completely filled and the diagonal matrix 
elements slightly deviate from the sum of 
energies of one-phonon components from which the complex configurations are 
built. The latter procedure is referred as the Pauli principle
correction. It leads to substantial increase of the computational time 
but its influence on the final results is usually marginal,
especially in the case of giant resonances.
The values of matrix elements of interaction between one- and three-phonon
configurations are on the level of the Pauli principle corrections and 
they are neglected in calculation presented in the following.

The advantage of the QPM approach with its pre-diagonalization of the model 
Hamiltonian on the QRPA level in comparison to the SRPA is that
it allows for an easier truncation of huge number of complex configurations 
before the matrix (\ref{QPM}) diagonalization. 
Indeed, the QRPA equations yield collective, weakly collective, and almost 
pure $1p1h$ solutions. The complex configurations which contain collective
phonons couple to doorway states much stronger then the ones made of only
non-collective phonons. In another words, the pre-diagonalization yields a
limited number of matrix elements $V_{1ph}^{2ph}$ which are substantially
larger than $v_{ijlk}$ in Eq.~(\ref{HA}) and, accordingly, play the most important role
in the damping mechanism.      
This will be discussed in detail below.

\section{Characteristic scales and giant resonance decay mechanisms}
\label{sec4}

\subsection{The ISGQR case: Coupling to low-lying surface vibrations}
\label{subsec41}

\begin{figure}
\begin{center}
\resizebox{0.48\textwidth}{!}{%
  \includegraphics{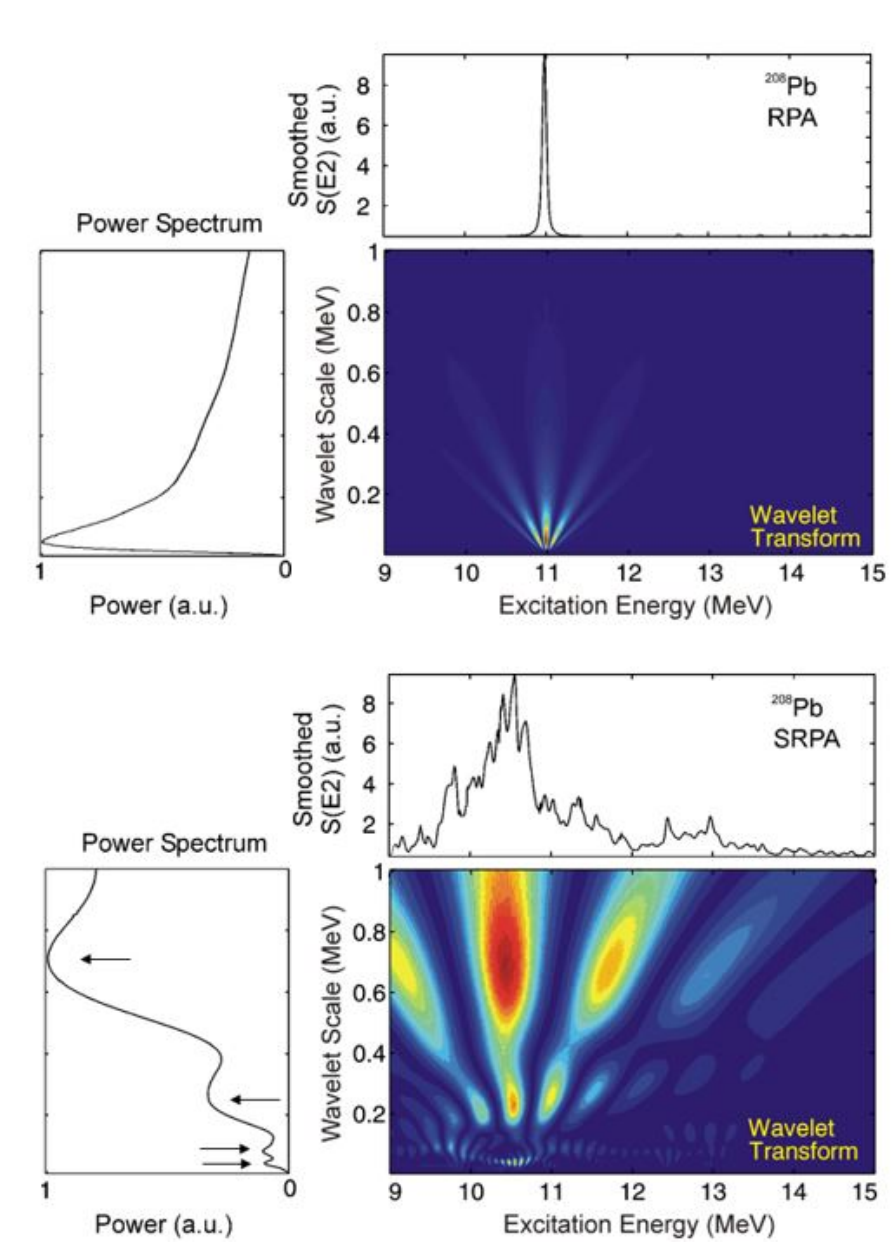}
}
\caption{
Same as fig.~\ref{fig221}, but for RPA (top) and SRPA (bottom) calculations of the E2 response in $^{208}$Pb.
}
\label{fig411}  
\end{center} 
\end{figure}

As explained above, the interpretation of characteristic  scales of the giant resonance fine structure requires a model comparison with microscopic calculations.
This is illustrated in fig.~\ref{fig411}, which presents a wavelet analysis of calculations of the ISGQR strength function in $^{208}$Pb analogous to the experimental data (fig.~\ref{fig221}).
Within the RPA models, where only $1p1h$ transitions are treated, the ISGQR strength in heavy nuclei is concentrated in a single state. 
Accordingly, as the upper part of fig.~\ref{fig411} demonstrates, the wavelet analysis does not detect any characteristic scales except a trivial scale from folding with a Gaussian of width of 50~keV (FWHM), put in to mimic the experimental resolution.

If one goes beyond the mean field approximation and includes the coupling to
$2p2h$ configurations, the ISGQR strength distribution in $^{208}$Pb shown in the lower part of fig.~\ref{fig411}  fragments into many states and fine structure appears. 
By way of example, the SRPA calculation of ref.~\cite{lac00} is shown here.
The wavelet transform and the power spectrum exhibit several characteristic scales. This fact is a demonstration of the significance of coupling to $2p2h$ configurations for the formation of fine structure and related characteristic scales, as found in the
experimental spectra. 
We remark that the maximum scale shown in fig.~\ref{fig411} is restricted to 1 MeV, as for the experimental data in figs.~\ref{fig221}, to achieve better visibility at small scales. 
A larger characteristic scale at 2.1 MeV representing the total width of the resonance is also found.

\begin{figure}
\begin{center}
\resizebox{0.48\textwidth}{!}{%
  \includegraphics{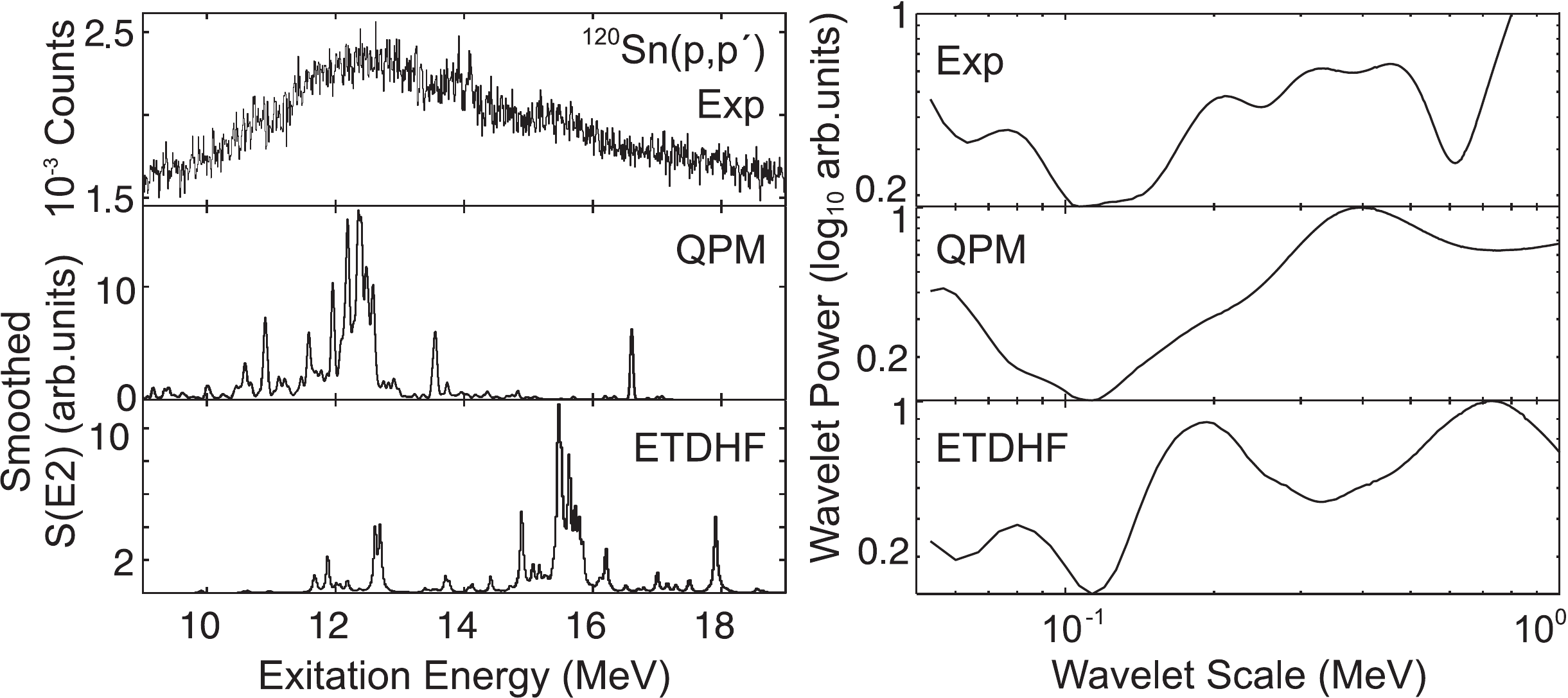}
}
\caption{
Left: Experimental spectrum of the $^{120}$Sn($p,p'$) reaction at the maximum of the ISGQR cross section versus QPM and ETDHF predictions for the ISGQR. 
Right: corresponding wavelet power spectra.
Figure taken from ref.~\cite{she09}.
}
\label{fig412}  
\end{center} 
\end{figure}

Because the models including $2p2h$ states work with different truncation schemes and based on different interactions, the characteristic scales obtained in the wavelet analysis also differ.
This has been discussed in detail for the case of $^{208}$Pb in ref.~\cite{she09}. 
Figure~\ref{fig412} presents $^{120}$Sn as another example, where on the left side the experimental spectrum (top) is compared with QPM (middle) and ETDHF (bottom) ISGQR strength distributions, and the right side contains the corresponding CWT power spectra. 
The QPM reproduces the experimental centroid energy while it is shifted to higher excitation energy by several MeV in the ETDHF results (however, this does not have an impact on the wavelet power). 
Both models fall short of describing the experimental widths. 
Correspondingly, the spectra contain little wavelet power at scales of several MeV in contrast to the data.  
The wavelet scale axes are again restricted to 1 MeV. 
Thus, a strong scale observed at several MeV in the experimental power spectrum is not visible. 
Overall, the QPM power spectra are closer to the data but show deviations from the experimental result.

The systematic investigation of the ISGQR fine structure from light to heavy nuclei led to the following conclusions:
(i) In spherical medium-mass and heavy nuclei the appearance of fine structure results from the coupling of $1p1h$ to $2p2h$ states, i.e.\ the doorway mechanism \cite{she04,she09}.
(ii) The models provide a semi-quantitative description of the experimental scales, where the number of scales in a certain scale energy region is reproduced rather than the exact values.
(iii) One always finds a large scale corresponding to the total width of the resonance.
The models typically find smaller widths, either due to the neglect of coupling to the continuum or due to the truncation of $2p2h$ model spaces.  
(iv) In lighter nuclei Landau fragmentation might become relevant.
For $^{40}$Ca, calculations with a modern realistic interaction derived from the unitary correlation operator method (UCOM) interaction find that the scales characterizing the fine structure are largely present already on the mean-field level. 
This finding is in contrast to many previous calculations of the ISGQR strength distribution in $^{40}$Ca, which all attribute the fragmentation to the spreading width (see ref.~\cite{usm11} and refs.\ therein).
(v) Deformation amplifies the role of Landau fragmentation for light \cite{usm16} and heavy \cite{kur18} nuclei.
(vi) Small scales are experimentally identified in light nuclei which cannot be explained by any of the calculations \cite{usm11,usm16}. 
These may result either from the coupling between $2p2h$ states neglected in the SRPA calculations or they may originate from Ericson fluctuations \cite{eri66}.

\begin{figure}
\begin{center}
\resizebox{0.35\textwidth}{!}{%
  \includegraphics{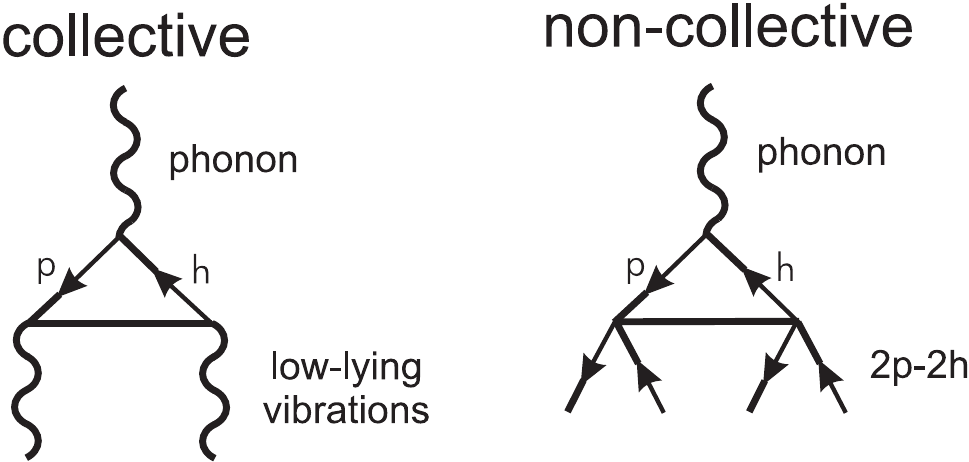}
}
\caption{
{\it Collective} vs.\ {\it non-collective} damping mechanisms. 
The term `collective' corresponds to coupling to low-lying surface vibrations \cite{ber83}. 
The non-collective contribution results from the mixing of initial $1p1h$ states with the large background of states with more complex wave functions.
Figure taken from ref.~\cite{she09}.
}
\label{fig413}  
\end{center} 
\end{figure}

While the quantitative results of the models differ, the success in reproducing at least qualitative features of the characteristic experimental scales motivated attempts to extract  their underlying physical nature from the model predictions. 
In the framework of the QPM calculations described in sec.~\ref{subsec32} a decomposition of the full model space into subspaces, corresponding approximately to
different damping mechanisms, is possible. 
One such important mechanism contributing to the damping of the single-particle \cite{gal88} as well as the collective response \cite{ber83} in heavy nuclei is the
coupling to low-lying surface vibrations.
The importance of this mechanism for the damping of giant resonances (in the following called \textit{collective} damping) was first discussed by Bertsch, Bortignon and Broglia  \cite{ber79,bro81,bor81}. 
Another significant contribution may come from mixing of the initial $1p1h$ states with the large background of incoherent $2p2h$ states, hitherto called \textit{non-collective} damping. 
These two mechanisms are depicted diagrammatically in fig.~\ref{fig413}.

\begin{figure}
\begin{center}
\resizebox{0.4\textwidth}{!}{%
  \includegraphics{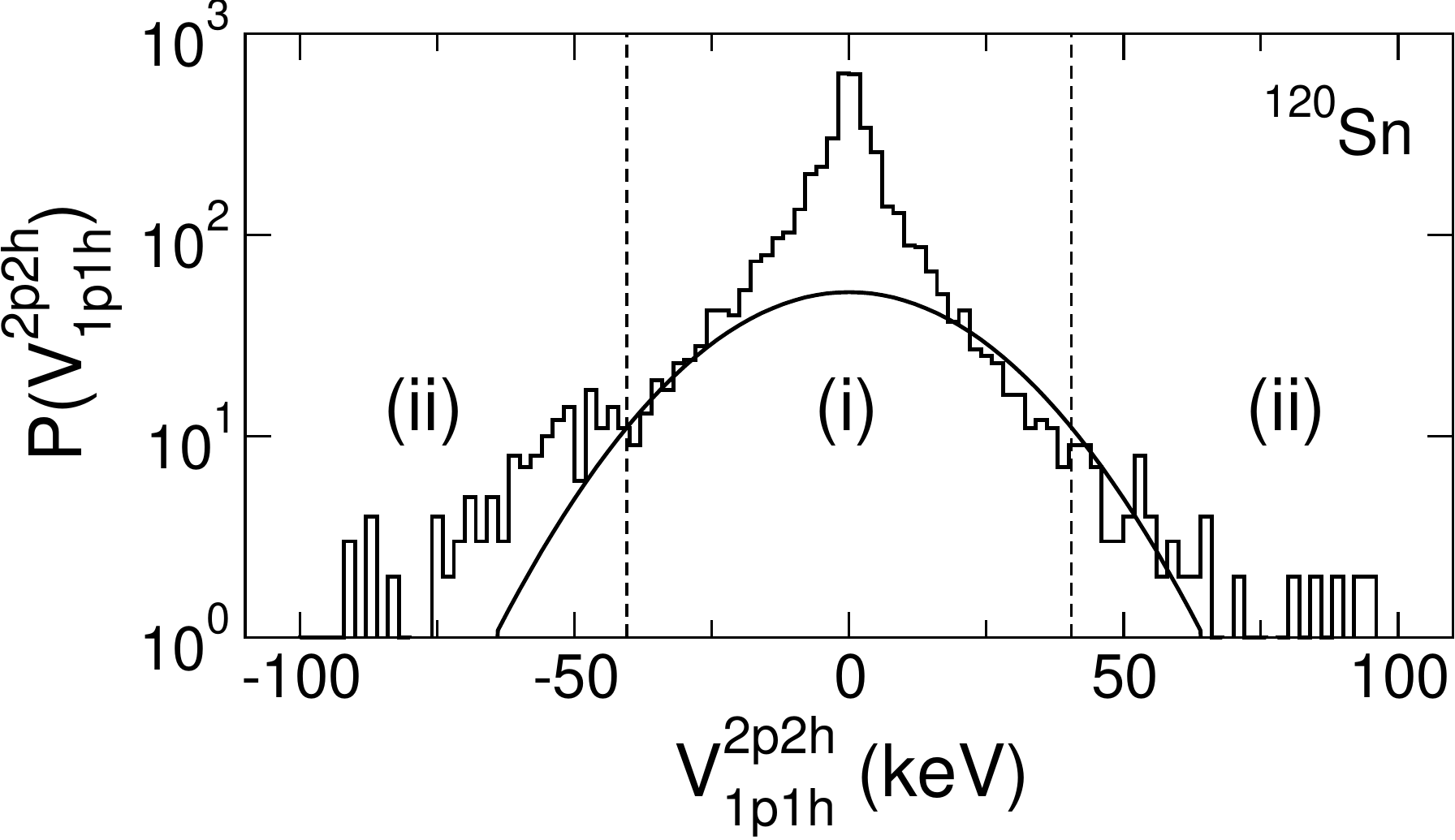}
}
\caption{
Distribution of the coupling matrix elements $V_{1p1h}^{2p2h}$ between $1p1h$ to $2p2h$ states in the QPM calculation for the ISGQR response in $^{120}$Sn. 
Figure taken from ref.~\cite{she09}.
}
\label{fig414}  
\end{center} 
\end{figure}

The two contributions can be approximately disentangled by investigating the properties of the coupling matrix elements between the between the $1p1h$ and $2p2h$, respectively 1- and 2-phonon configurations (in the language of the QPM).
The probability $P$ of finding a certain value of the coupling matrix $V^{\rm 2ph}_{\rm 1ph}$ in the QPM is displayed as a histogram in fig.~\ref{fig414} for the case of $^{120}$Sn as an example.
The solid line shows a Gaussian distribution expected for fully chaotic systems from the Gaussian orthogonal ensemble (GOE) as predicted by Random-Matrix-Theory (RMT) \cite{guh98}. 
The value of the Gaussian width is adopted to match the data.
The distribution deviates appreciably from the Gaussian form: one finds a strong overshoot of very small matrix elements and some enhancement at large values. 
Similar features have been reported from the analysis of off-diagonal interaction matrix elements in shell-model calculations \cite{zel96}.
The excess of small matrix elements indicates that many of the two-phonon configurations contribute little to the fragmentation process. 
On the other hand, the large matrix elements have an appreciable effect and are due to the presence of soft collective modes.
An approximate separation of transitions contributing to collective and non-collective damping can be achieved by assigning subspaces and repeating the diagonalization within these subspaces. 
This is indicated in fig.~\ref{fig414} as (i) and (ii) for non-collective and collective damping, respectively. 

%
\begin{figure}[b]
\begin{center}
\resizebox{0.48\textwidth}{!}{%
  \includegraphics{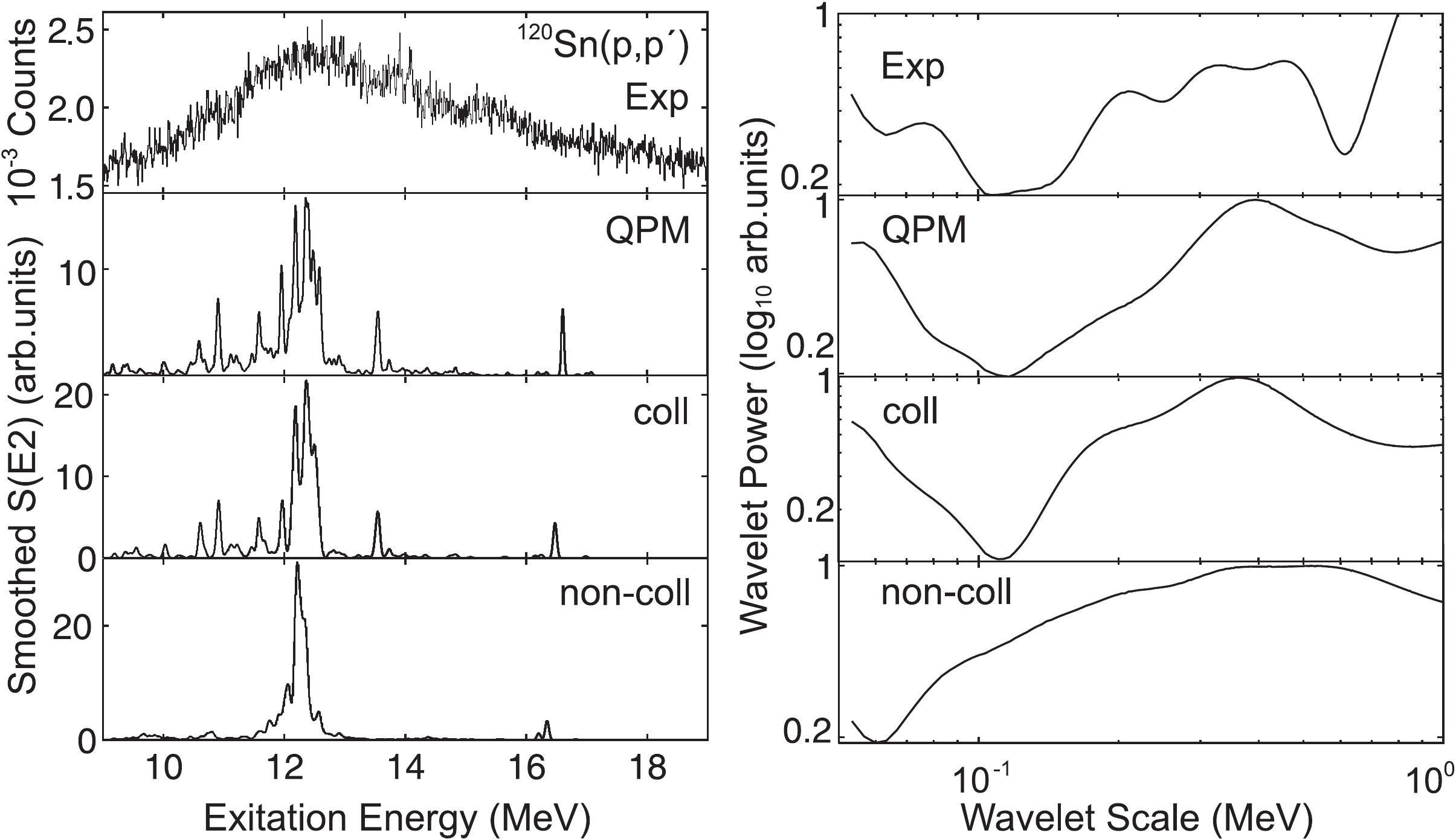}
}
\caption{
Left: Experimental spectrum of $^{120}$Sn as compared to the QPM prediction for E2 strength and its decomposition into the collective and non-collective damping contributions. 
Right: corresponding wavelet power spectra.
Figure taken from ref.~\cite{she09}.
}
\label{fig415}  
\end{center} 
\end{figure}
The resulting total (cf.~fig.~\ref{fig412}), collective and non-collective ISGQR strength functions are displayed in the left panel of fig.~\ref{fig415}.
In both cases it is obvious that the fragmentation is dominated by the collective mechanism.
However, one should be aware that the full calculation is not just
the sum of the two contributions, and interference terms may play a
role.
The corresponding wavelet power spectra displayed in the right panel of fig.~\ref{fig415} clearly demonstrate that all scales are already present in the collective part. 
The non-collective part shows a wavelet power distribution broadly distributed over the range of scales.
Similar results are found for other nuclei and models \cite{she09}.

\begin{figure}
\begin{center}
\resizebox{0.5\textwidth}{!}{%
  \includegraphics{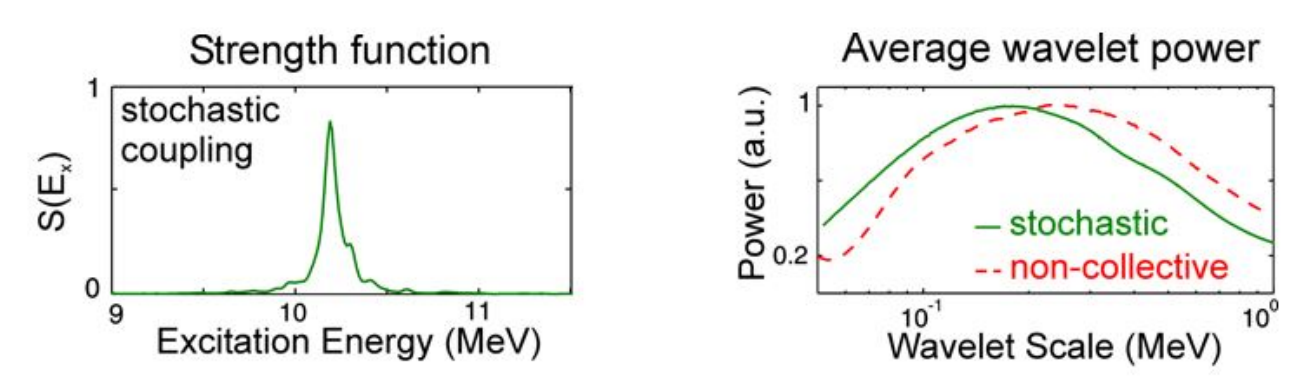}
}
\caption{
Strength function and wavelet power from a stochastic coupling model for the ISGQR in $^{120}$Sn.
}
\label{fig416}  
\end{center} 
\end{figure}
The absence of pronounced scales in the non-collective part suggests a generic origin, i.e., a stochastic coupling to the background of complex states. 
Then, the level spacings and coupling matrix element distributions should be given by the GOE. In order to test this assumption, we have generated a GOE with the same Gaussian distribution of the matrix elements as found in the QPM calculation of $^{120}$Sn. 
After averaging over a sufficient ensemble of random copies one obtains the strength function shown in the left part of fig.~\ref{fig416}.
Wavelet analysis leads to the power spectrum displayed in the right part of fig.~\ref{fig416} as the dashed line. 
Indeed, the stochastic coupling model produces a large variety of scales which manifests itself in a broad distribution, exactly what is seen in the QPM results in in fig.~\ref{fig414} for the non-collective subspace (i). 
The slight shift of the maximum compared to the stochastic model can probably be traced back to the decomposition procedure. 
Rather than taking all matrix elements exceeding the Gaussian distribution at a given strength,
one has to define a cut-off value indicted by the dashed lines in fig.~\ref{fig414} below which all matrix elements are assumed to belong to subspace (i).
As demonstrated in ref.~\cite{she09}, collective coupling dominates the spreading for heavy nuclei, but the non-collective coupling mechanism becomes increasingly important for lighter nuclei.

\subsection{The IVGDR case: Landau damping}
\label{subsec42}

Although fine structure of the IVGDR is now systematically established on a level comparable to the ISGQR, we focus in the discussion on the $^{208}$Pb and $^{120}$Sn studied in more detail in sec.~\ref{subsec41}.
Some general conclusions are discussed at the end. 

\begin{figure}
\begin{center}
\resizebox{0.48\textwidth}{!}{%
  \includegraphics{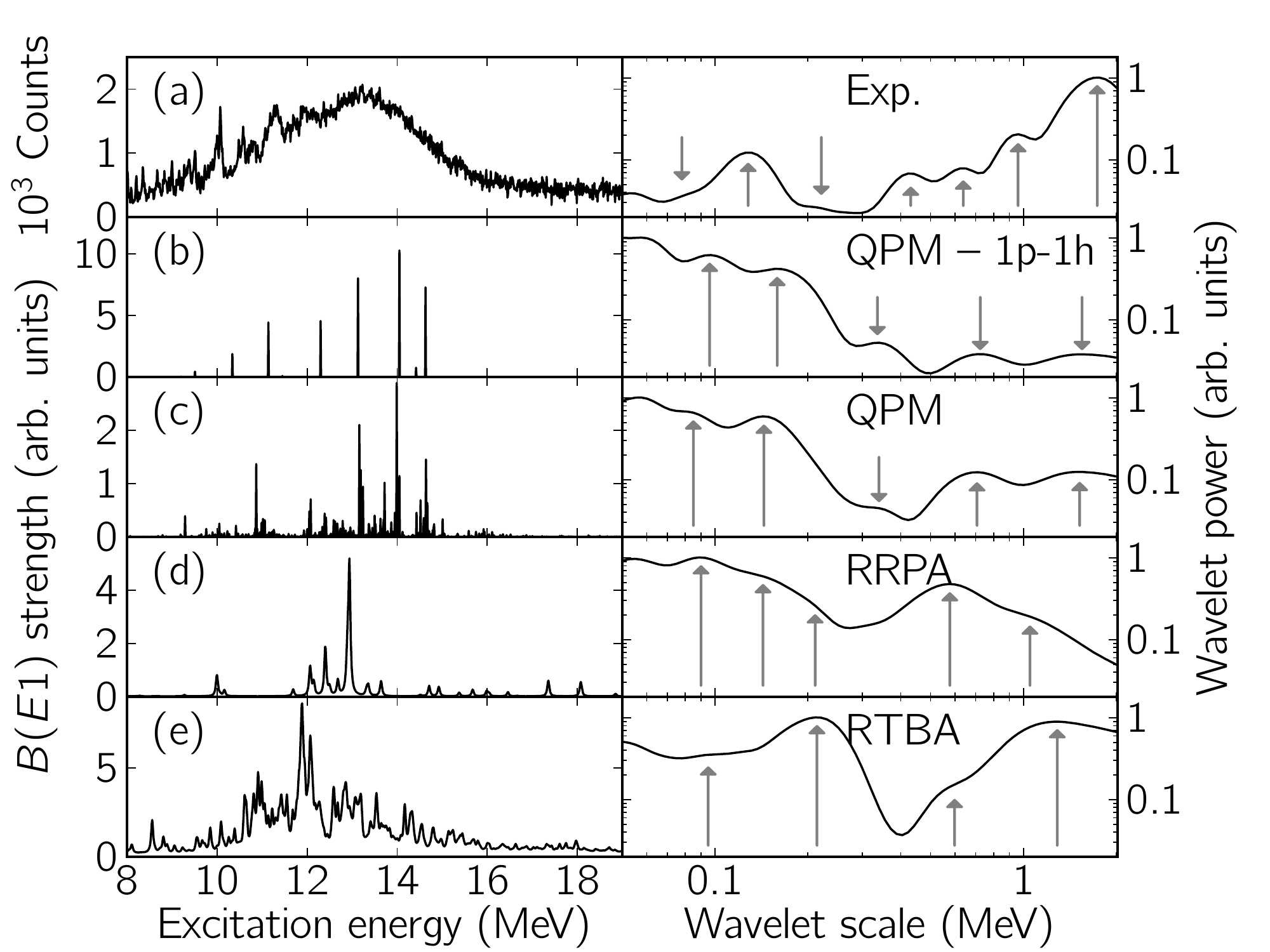}
}
\caption{
(a) Experimental spectrum of the IVGDR studied with the $^{208}$Pb(p,p$^\prime$) reaction in comparison with theoretical predictions of  the B(E1) strength distribution in $^{208}$Pb (left side) and the resulting power spectra from a CWT analysis (right side). 
Theoretical results are shown for the QPM with 1-phonon (b) and (1+2)-phonon (c) model spaces, RRPA (d)  and RTBA (e). 
Characteristic scales are marked by arrows.
Figure taken from ref.~\cite{pol14}.
}
\label{fig421}  
\end{center} 
\end{figure}
The pronounced fine structure found in $^{208}$Pb (cf.~fig.~\ref{fig213})  has been analyzed by QPM \cite{rye02} and RTBA \cite{lit07} calculations, the latter based on a relativistic mean-field approach.
The comparison of the experimental spectrum (left side) and the wavelet power spectrum (right side) resulting from the CWT analysis with those of the model calculations for the B(E1) strength distributions is shown in fig.~\ref{fig421}. 
It should be noted that the experimental spectrum, (a), does not represent the B(E1) strength but rather the Coulomb excitation cross section, which is modified by the excitation-energy dependent virtual photon number \cite{vnc19}. 
Extraction of the experimental B(E1) distribution is possible \cite{tam11,has15,bir17,mar17}. 
However, the need to disentangle the E1 cross section from other contributions can only be achieved for larger energy bins, where the information on the fine structure is partially lost. 
Such a conversion of the experimental data to B(E1) strength would lead to a slight
energy shift ($< 5$\%) of the characteristic scales and an increase of relative power towards higher excitation energies.

A QPM calculation on the 1-phonon level (b) results in a B(E1) strength distribution dominated by five transitions distributed between 11 and 15 MeV with a centroid
energy of 13.25 MeV (defined as $m_1/m_0$, where $m_i$ denotes the $i$th moment of the distribution). 
The experimental centroid energy of 13.43 MeV is fairly well reproduced. 
Inclusion of 2-phonon configurations, fig.~\ref{fig421}(c), leads to fragmentation but the dominant $1p1h$ transitions remain and the centroid energy is unaffected.
A similar comparison of the RRPA and RTBA results, fig.~\ref{fig421}(d) and (e), shows somewhat larger differences of the distributions although the centroid energy is hardly affected.

Since there is no absolute scale, the corresponding CWT power spectra shown on the right side of fig.~\ref{fig421} are normalized relative to each other. 
Overall, both models broadly reproduce the variation of power with scale value. 
A power peak at small scales around 100 to 200 keV is followed by a minimum of power at a few hundred keV and another rise towards larger values. 
The scale values of power maxima and minima are better reproduced by the QPM. However, the relative ratio of maxima at smaller and larger scales is predicted to decrease in the QPM while experiment shows an increase. 
In the RTBA the ratio is closer to the data. 

The comparison of  figs.~\ref{fig421}(b,c) and (d,e) provides information on the damping mechanism responsible for the fine structure. 
Clearly, the QPM results show structure already at the 1-phonon level. 
The appearance of scales $>$ 1 MeV can be easily understood by the spacing of the five dominant transitions, but the wavelet analysis of the 1-phonon result also finds the characteristic scales with smaller values $<$ 1 MeV. 
The similarity between the power spectra and scales deduced from the QPM calculation
for 1-phonon (b) and (1+2)-phonon (c) states suggests that the fragmentation of $1p1h$ transitions (i.e., Landau damping) is the most important mechanism leading to fine structure of the IVGDR in $^{208}$Pb. 
The coupling to complex configurations and, in particular, to low-lying collective vibrations identified as dominant mechanism for the ISGQR seems to play a minor role only for the generation of the fine structure.
While the relative weight changes, major scales are also found at about the same energies in the CWT analysis of the RRPA (d) and  and RTBA (e) results. 
The observation of characteristic scales in the RRPA calculation again supports
the above conclusion.

\begin{figure}
\begin{center}
\resizebox{0.4\textwidth}{!}{%
  \includegraphics{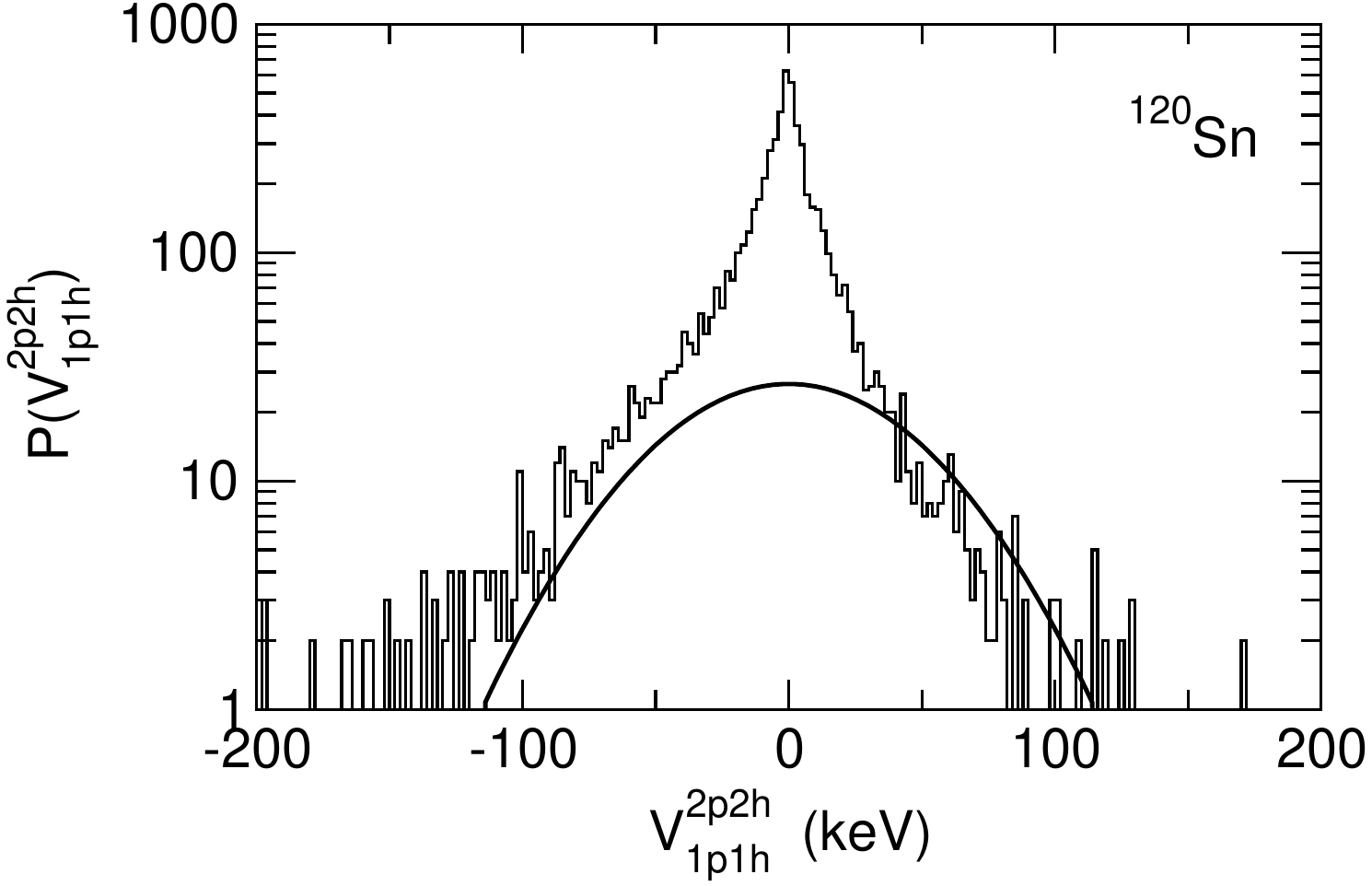}
}
\caption{
Same as fig.~\ref{fig414}, but for the IVGDR response in $^{120}$Sn.
}
\label{fig422}  
\end{center} 
\end{figure}
We have also performed an analysis of the coupling matrix elements $V^{\rm 2ph}_{\rm 1ph}$ in analogy to the ISGQR case, again for $^{120}$Sn as a representative example.
The probability distribution in the case of the IVGDR is depicted in fig.~\ref{fig422}.
A similar excess of small coupling matrix elements compared to the RMT prediction (solid line) as in the case of the ISGQR is observed.
However, these do not contribute to the fragmentation process.
At large coupling strength the QPM result is much closer to the RMT limit.

\begin{figure}
\begin{center}
\resizebox{0.3\textwidth}{!}{%
  \includegraphics{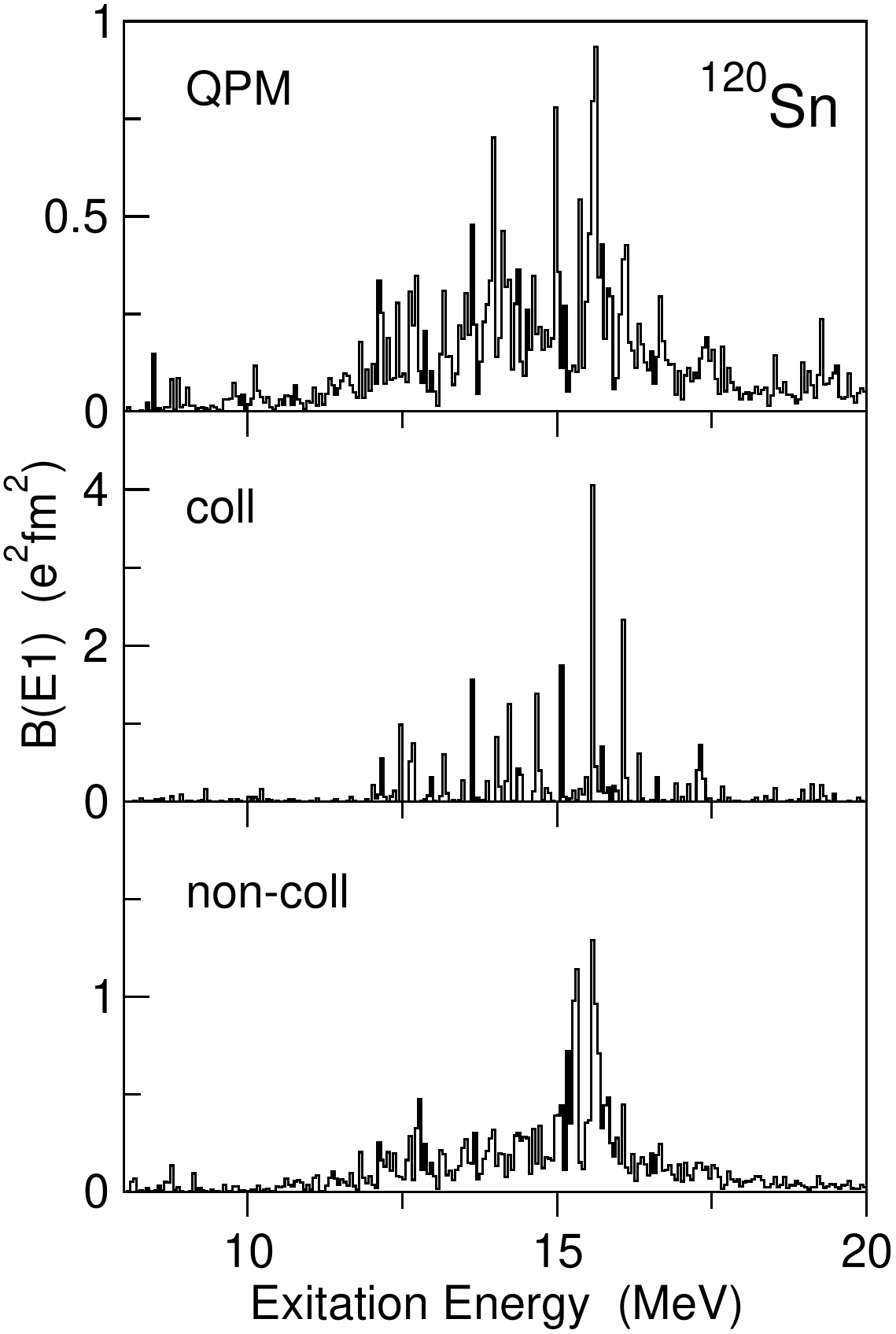}
}
\caption{
Same as the left side of fig.~\ref{fig415}, but for the IVGDR response in $^{120}$Sn.}
\label{fig423}  
\end{center} 
\end{figure}

The E1 strength distribution of $^{120}$Sn resulting from the full QPM calculations is displays in the top part of fig.~\ref{fig423}.
When the QPM model space is artificially divided into subspaces corresponding to the value of $V^{\rm 2ph}_{\rm 1ph}$ as indicated in by the dashed lines in fig.~\ref{fig414}, strength distributions from the larger (collective) and smaller (non-collective) matrix elements shown in the middle and bottom part of fig.~\ref{fig423}, respectively, exhibit comparable fragmentation.
Thus, the relevance of damping through the coupling to low-lying surface vibrations for the case of the IVGDR cannot be discovered by such an analysis.

The findings on the fine structure of the IVGDR can be summarized as follows:
(i) The systematic investigation of medium-mass to heavy nuclei corroborates the importance of Landau damping as the main source of the IVGDR fine structure \cite{jin18}.
(ii) The quantitative description of the experimental scales by the different models is comparable to the ISGQR case, i.e.\ the models reproduce the number of scales in a certain scale energy region rather than exact values.
(iii) In some cases small characteristic scales have been found, which could not be explained by calculations on the RPA level \cite{fea18,jin18}.
These are likely to result from the spreading width although at present we cannot distinguish whether these are generated by the collective or the non-collective mechanism discussed above.   
(iv) Fine structure is also observed in heavy deformed nuclei \cite{don18} despite the extremely high level densities.
(v) In light deformed nuclei, the comparison between the experimental data and 
RPA calculations based on modern realistic interactions suggest that fine structure at the level of a few hundred keV results mainly from the deformation of the nuclei driven by α clustering \cite{fea18}.
These data also show in all nuclei studied a small scale not explained by any calculation.
Since the values are consistent with expectations of the average width due to Ericson fluctuations in this energy region \cite{eri66,ebe72}, we tentatively follow this identification.

\section{Conclusions and outlook}
\label{sec5}

In recent years fine structure has been observed as a global phenomenon in spectra of high-resolution experiments for all types of giant resonances.
In particular, the ISGQR and IVGDR have been studied systematically across the nuclear chart and with respect to the role of deformation. 
Wavelet analysis has been developed as a method to extract quantitative information on the fine structure in terms of characteristic scales.
The corresponding analysis of strength distributions from theoretical approaches incorporating one or several mechanisms of giant resonance decay permits to identify the origin of the observed fine structure.

By way of example, we discussed in detail two approaches, SRPA and QPM, widely used for such an analysis which represent different approximations to the inclusion of the doorway state mechanism.
Indeed, spreading through the first step of the doorway mechanism, i.e.\ coupling between one particle-one hole ($1p1h$) and two particle-two hole ($2p2h$) states is identified as generator of the fine structure of the ISGQR.
In heavy nuclei it is dominated by coupling to low-lying surface vibrations, a mechanism first investigated by Bertsch, Bortignon, Broglia and Dasso \cite{ber79}. 
In lighter nuclei stochastic coupling becomes increasingly important.
In contrast, the fine structure observed for the IVGDR arises mainly from the fragmentation of the $1p1h$ strength, i.e.\  Landau damping, although some indications for the relevance of the spreading width are also found.  

Concerning future work, the systematic observation of fine structure of the ISGQR and the IVGDR immediately raises the question about similar observations for the isoscalar giant monopole resonance (ISGMR).
An extensive study at iThemba LABS using $0^\circ$ $\alpha$ scattering indeed confirms the fine structure phenomenon for the ISGMR as well \cite{vnc19} and  a comprehensive analysis is underway.
Finally, we would like to mention recent initiatives to utilize giant resonance $\gamma$ decay as an alternative probe.
Although a weak branch, it carries unique information as discussed by Bortignon and coworkers  \cite{bor84,bre12}. 
Pioneering experiments reported the extraction of total decay branching ratios from the  IVGDR and the ISGQR to low-lying states \cite{bee89,bee90a,bee90b,pon92}.
With the advent of new large-volume LaBr detectors with much improved $\gamma$ and time resolution \cite{gia13} and in combination with high-resolution spectrometers one can hope in future experiments to `scan' the $\gamma$ decay across the resonances \cite{vnc19}. 

\begin{acknowledgement}

This work was funded by the Deutsche Forschungsgemeinschaft (DFG, German
Research Foundation) -- Projektnummer 279384907 -- SFB 1245.
      
\end{acknowledgement}

\end{document}